\documentclass[%
 reprint,
 amsmath,amssymb,
 aps,
]{revtex4-1}

\usepackage{graphicx}
\usepackage{dcolumn}
\usepackage{xcolor}
\usepackage{bm}

\def\be{\begin{equation}}
\def\ee{\end{equation}}
\def\bmu{\begin{multline}}
\def\bea{\begin{eqnarray}}
\def\eea{\end{eqnarray}}

\def\f{\frac}

    \makeatletter
    \renewcommand*{\@fnsymbol}[1]{\ensuremath{\ifcase#1\or *\or \dagger\or \ddagger\or
        \mathsection\or \mathparagraph\or \|\or **\or \dagger\dagger
        \or \ddagger\ddagger \else\@ctrerr\fi}}
    \makeatother

\begin{document}

\title{A Time-Dependent Evolutionary Strategy to Evolve Generalists}
\title{Tuning Dynamic Environments to Evolve Generalists}
\title{Tuning Time-Dependent evolutionary strategies to evolve and maintain generalists}
\title{Resonant environmental cycling to evolve and maintain generalists}
\title{Tuning environmental timescales to evolve and maintain generalists}
%

\author{Vedant Sachdeva$^{1\dagger}$, Kabir Husain$^{2\dagger}$, Jiming Sheng$^3$, Shenshen Wang$^{3\ast}$, Arvind Murugan$^2$}
\email{Correspondence: shenshen@physics.ucla.edu, amurugan@uchicago.edu \\ $\dagger$ These authors contributed equally}
\affiliation{$^1$Graduate Program in Biophysical Sciences, University of Chicago, Chicago, IL, $^2$Department of Physics, University of Chicago, IL, $^3$Department of Physics and Astronomy, University of California Los Angeles, Los Angeles, CA}







\begin{abstract}

Natural environments can present diverse challenges, but some genotypes remain fit across many environments. Such `generalists' can be hard to evolve, out-competed by specialists fitter in any particular environment. Here, inspired by the search for broadly-neutralising antibodies during B-cell affinity maturation, we demonstrate that environmental changes on an intermediate timescale can reliably evolve generalists, even when faster or slower environmental changes are unable to do so. We find that changing environments on timescales comparable to evolutionary transients in a population enhances the rate of evolving generalists from specialists, without enhancing the reverse process. The yield of generalists is further increased in more complex dynamic environments, such as a `chirp' of increasing frequency. Our work offers design principles for how non-equilibrium fitness `seascapes' can dynamically funnel populations to genotypes unobtainable in static environments.
\end{abstract}
\maketitle

\footnotetext{Correspondence: shenshen@physics.ucla.edu, amurugan@uchicago.edu}




%



%

%
%
%
%

Evolutionary outcomes are driven by environmental pressures, but environments are rarely static\cite{Levins1968-qz}. In a changing environment, some genotypes -- termed generalists -- maintain a uniformly high fitness over time, even if they are not globally fit at any particular instant. A striking example is that of broadly-neutralizing antibodies (bnAbs) against HIV and other viruses -- these antibodies maintain potency against the large diversity of viral strains that may arise in an infected individual over time \cite{Burton2004-jw,Burton2012-zd,Wu2010-xh}. It is desirable for the immune system to select for generalist antibodies during B-cell affinity maturation, a rapid evolutionary process\cite{Cobey_Sarah2015-tb}, but generalists are often out-competed by specialists that only bind particular viral strains. 


Recent work has suggested that sequential vaccination with different viral antigens, rather than a single cocktail of those antigens, can better select for generalist antibodies during affinity maturation \cite{Pissani2012-to, Malherbe2011-cg,Wang2017-tx,Wang2015-jg}. This result is consistent with the broader idea that a time-varying environment can drive evolution out of equilibrium and into genotypes unevolvable in static environments \cite{Mustonen2010-kf,Mustonen2009-fu,Arndt2004-yq,Kussell2014-mg,Goldenfeld2011-pm}. However, the broader principles underlying generalist selection by dynamic environments remain unknown. In particular, the interplay of environmental and evolutionary timescales and choices of correlated antigens generates a high-dimensional space of possible vaccination protocols. Hence guiding principles are needed to find optimal protocols for evolving generalist genotypes.

Here, we take a phenomenological approach to design dynamic environments that select generalists. We analyze situations in which generalists are entropically disfavoured or isolated by fitness valleys, and thus unevolvable in a static environment. We find that a dynamic environmental protocol can maximize the yield of generalists if the environment changes on the same timescale as the evolutionary transients of the population, i.e., on the timescale for allele frequencies to reach steady state. Consequently, switching antigens before antibody populations have evolved to a steady state can dynamically funnel finite populations from specialists to generalists, even when faster or slower switching is unable to do so.

We understand these results in terms of a kinetic asymmetry between generalists and specialists. Environmental dynamics at the right timescale perturbs specialist populations while leaving generalists relatively undisturbed. This asymmetry favours evolution from specialists to generalists without enhancing the time-reversed process. 
In contrast, faster or slower environmental dynamics may be cast into effective static fitness landscapes \cite{Cvijovic2015-xy}, and are thus unable to maintain a strong kinetic asymmetry between specialists and generalists. In this sense, the intermediate cycling mechanism studied here exploits a truly non-equilibrium evolutionary `seascape' \cite{Kussell2014-mg,Mustonen2009-fu} with no static analog. 

Our framework proposes novel protocols for evolving generalists, such as a `chirp' where the environment is cycled at an increasing frequency, and predicts optimal correlations between antigens to be used. Since we use a sufficiently abstracted model of B-cell affinity maturation,  our analysis might be adapted for other temporal evolution protocols, e.g., to avoid antibiotic resistance\cite{Toprak2011-ae,Marrec2018-vv,De_Jong2018-ao} and for cancer treatments \cite{Gatenby2009-hp,Katouli2011-uk}. 

Numerous works have studied evolution in time-varying environments, including in the context of evolving generalists \cite{Kassen2002-co,Desponds2016-ko,Uecker2011-kn,Hemery2015-hv,Kashtan2005-zp,Lipson2002-un,Xue2016-al,Raman2016-bj}. Relatively fewer works\cite{Cvijovic2015-xy, Mustonen2008-im,  Kussell2006-ch, Mayer2017-gl} have analyzed the case of intermediate timescales where the environment changes before populations reach steady state. In this broader sense, our work is a step towards a theory of evolution in time-varying environments with no separation of timescale between the evolutionary response of populations and environmental changes.



\begin{figure}
\begin{centering}
\includegraphics[width=0.5\textwidth]{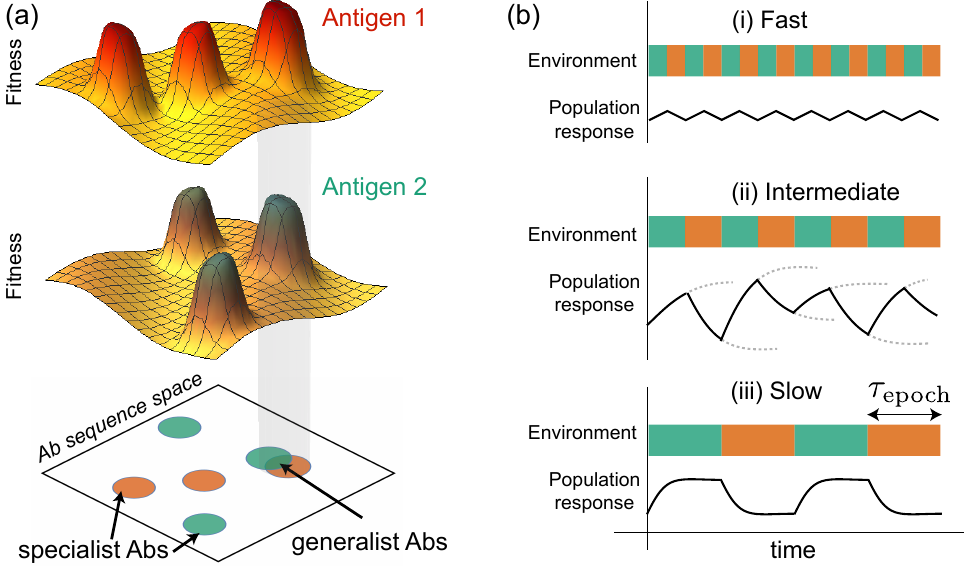}
\caption{Time-varying environments on intermediate timescales can dynamically funnel specialists to generalists. (a) Generalist antibodies that bind multiple antigens can be hard to evolve during B-cell affinity maturation as compared to specialists that only bind one antigen. Specialists for an antigen can constitute a single (Fig.\ref{fig:hisifreq}) or multiple islands (Fig.\ref{fig:Jijmodel}) in antibody sequence space. (b) We consider time-varying selection pressure on timescales (i) fast, (ii) intermediate or (iii) slow relative to evolutionary transients. In the intermediate regime, the selection pressure (e.g., antigen) changes before evolutionary transients (dashed lines) are complete and a steady state is reached.  
\label{fig:schematic}}
\end{centering}
\end{figure}

\section*{Results}

We present our results in two broad models of how specialists and generalists can be distributed in sequence space, paralleling different assumptions about antigen-antibody binding. {In both cases, we model populations (e.g., the population of B-cells across all germinal centers in an organism). We explain our results in terms of the rate at which a population of specialists evolves generalists in time-varying environments relative to the rate of the time-reversed processes from generalists to specialists}.

\begin{figure*}
\begin{centering}
\includegraphics{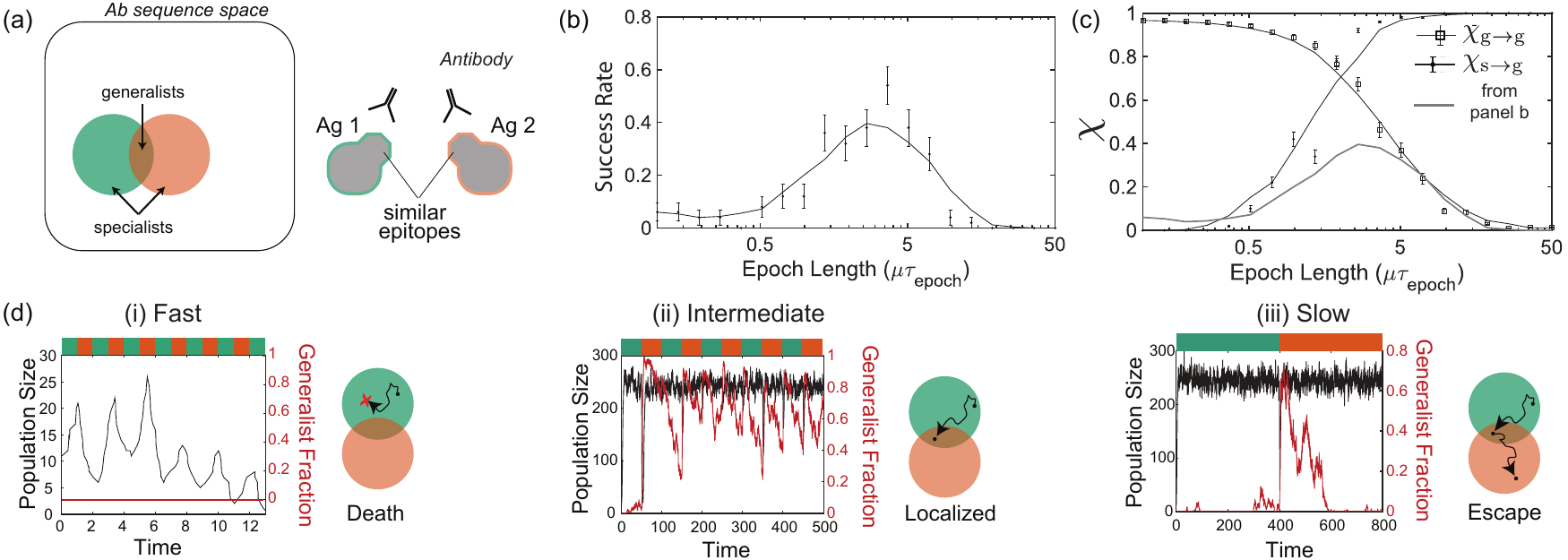}
\caption{
Intermediate timescale cycling of antigens strikes a balance between evolving and maintaining rare generalist antibodies. 
(a) We assume many specialist antibodies can bind each antigen at a partially conserved epitope; see text for model. Generalists and specialists have similar fitness.
(b) Cycling antigens at an intermediate timescale $\tau_{\text{epoch}}$ most reliably yields generalists in repeated $K=500$ population simulations.   
(c) An initially-specialist population is more likely to evolve generalists (higher $\chi_{\text{s}\to\text{g}}$) with slower cycling since (d,i) fast cycling typically leads to death of the entire population before any generalists are evolved. In contrast, slow cycling allows generalists to specialize; the fraction $\chi_{\text{g}\to\text{g}}$ of an initially-generalist population  that remains generalists falls with $\tau_{\text{epoch}}$ (see (d,iii)). 
(d,ii) Intermediate timescale switching allows sufficient time for generalists to evolve from specialists without providing enough time for generalists to specialize. 
\label{fig:hisifreq}}
\end{centering}
\end{figure*}

\subsection{Entropically disfavored generalists}
A basic difficulty in evolving generalists is that generalists are often far fewer in number than specialists. This is schematically shown in Fig.~\ref{fig:hisifreq}a, where specialists in each environment form a connected set of genotypes of similar fitness. The relatively few generalists, found at the intersection of such sets, can easily mutate into the more numerous specialists in any fixed environment.

We study the problem quantitatively in a simplified molecular model of antigen-antibody binding, as used for affinity maturation against HIV antigens. Antibodies bind to a single epitope, partially conserved across antigens $\eta = 1$, $2$. A (binary) antibody sequence $\mathbf{x}$ binds to an epitope sequence $\mathbf{h}^{\eta}$ with an affinity given by an additive sum-of-sites model: $\mathbf{x}\cdot \mathbf{h}^{\eta}$. Antibodies that bind above a threshold $T$ are assigned fitness $s (\epsilon -1) > 0$, while those that bind weaker have fitness $-s < 0$. We take $1 < \epsilon < 2$, such that the average fitness of an antibody across antigens is negative.

Since the epitope is relatively but not entirely, conserved across antigens, $\mathbf{h}^\eta$ for different antigens are assumed to share a conserved region of length $L_{c} = 12$ but have a variable region of length $L_{v}= 7$ \cite{Wang2015-jg} (see SI for other choices).  While based on a simple model of molecular binding, our results below apply broadly to the phenomenological description of specialists as connected islands of relatively uniform fitness, with no fitness barriers separating the generalists.

We simulate a finite population ($N \sim 500$) of antibodies in an environment that switches between antigens 1 and 2 on a timescale $\tau_{\text{epoch}}$ using a birth-death model (see SI), working in the limit of frequent mutations ($\mu N > 1$). Initializing a monoclonal population in a random specialist state for antigen $\eta = 1$, we monitor the fraction of generalists in the populations at late times (Fig. \ref{fig:hisifreq}d), systematically varying the timescale of switching $\tau_{\text{epoch}}$. Averaging over many simulation, we find that neither fast nor slow cycling is able to reliably elicit generalists in the population; however, an intermediate timescale of switching is able to do so (Fig.\ref{fig:hisifreq}b).


We sought to understand the origin of this non-monotonic behaviour by examining population dynamics in the limits of fast and slow cycling. For fast cycling, (i.e., small $\tau_{\text{epoch}}$), the initial specialist population is repeatedly confronted with an antigen it cannot bind to. Without enough time to mutate into a generalist, purifying selection drives the population to extinction (Fig. \ref{fig:hisifreq}d,i). Consequently, the fraction of trials in which specialists evolve into generalists, $\chi_{\text{s} \to \text{g}}$, is low (Fig. \ref{fig:hisifreq}c). 

In fact, in this limit the dynamics of the population are effectively described by a static, average landscape, where the specialist has fitness $s (\epsilon - 2) < 0$. In this regime, we find that purifying selection drives the population to extinction when $s > \mu \log N $; see SI for derivation and discussion of alternative cases. 

\begin{figure*}
\begin{centering}
\includegraphics{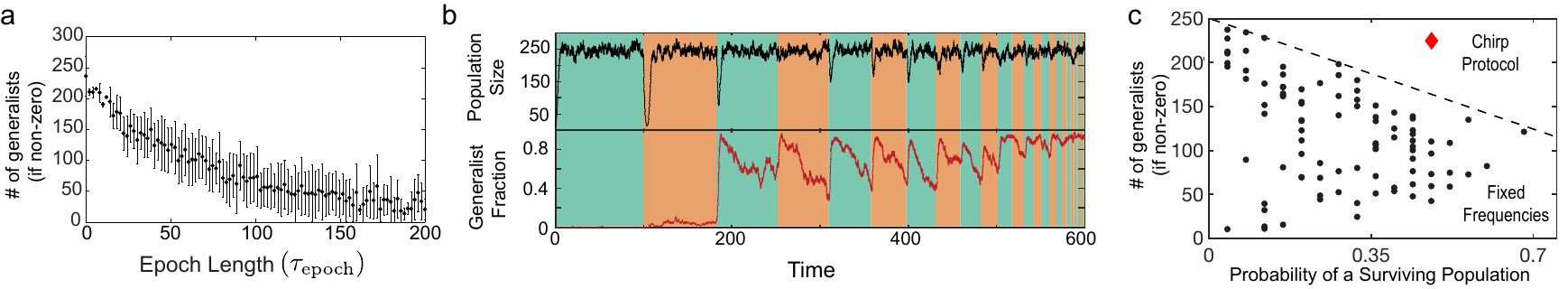}
\caption{Chirped cycling yields larger generalist populations more robustly than fixed frequency cycling. (a) While fast cycling usually results in population death, if the population does survive, the size of the generalist population produced is largest in this limit. (b,c) Thus fixed frequency cycling faces a trade-off between the probability of survival and the size of the generalist population, if the population does survive.  A chirped protocol, where the cycling frequency is increased over time, $\tau_{\text{epoch}} \rightarrow \frac{5}{6}\tau_{\text{epoch}}$ after each epoch, beats this trade off. Slow cycling early on allows the population to survive and evolve generalists; subsequent fast cycling maintains a large generalist population. \label{fig:chirp}}
\end{centering}
\end{figure*}


On the other hand, for very slow cycling (large $\tau_{\text{epoch}}$), any generalists that arise have enough time to specialize again by mutational drift (Fig. \ref{fig:hisifreq}d,iii). As a result, the fraction of an initially-generalist population that stay generalists over an environmental cycle, $\chi_{\text{g}\to\text{g}}$, falls with $\tau_{\text{epoch}}$, as seen in Fig.\ref{fig:hisifreq}c.

Consequently, we find that intermediate timescale cycling strikes a balance: providing enough time to for specialists to evolve into generalists (high $\chi_{\text{s} \to \text{g}}$), but not enough time for generalists to switch back to specialists again (high $\chi_{\text{g} \to \text{g}}$). In the SI, we determine this regime to be,
\begin{equation}
\tau_{\text{min}} \sim \frac{1}{\mu} C_{\text{init}}(L_c,L)  <   \tau_{\text{epoch}} <   \tau_{\text{max}} \sim \frac{1}{\mu} D(L_c,L) \log N
\end{equation} 
where $C_{\text{init}}$ and $D$ are combinatorial factors that account for the mutational distance of the initial naive repertoire from generalists and the number of generalist genotypes, respectively; see SI.

Note that if population sizes are small and sequence space is large, we find $\tau_{\text{min}} > \tau_{\text{max}} $, i.e., it takes longer for generalists to evolve from specialists than to specialize again. In this regime, the entropic bias in sequence space driving generalists to specialists is large; even fixed frequency cycling may not produce generalists.

Hence, we propose a new dynamic protocol - a `chirp' - that can alleviate this tension between evolving generalists from specialists ($\chi_{\text{s} \to \text{g}}$), which requires slower cycling, and the ability to maintain a population of generalists ($\chi_{\text{g} \to \text{g}}$), which requires faster cycling. A chirp, shown in Fig.~\ref{fig:chirp}, starts with slow cycling and increases the cycling frequency over time. Such highly dynamic `chirp' protocols outperform any fixed frequency cycling protocol; see Fig.~\ref{fig:chirp}c.

\subsection{Generalists isolated by fitness valleys}
We now consider a more general case where fitness valleys separate viable genotypes, and specialists and generalists form disconnected sets in sequence space. Such models have been used to describe antibodies for influenza and malaria \cite{Munoz2005-am,Chaudhury2014-tq, Deem2003-yu,Childs_Lauren_M2015-zn}, as well as describing RNA molecular fitness landscapes \cite{Pressman2019-at, Blanco2019-to}. In the affinity maturation context, such a model naturally arises if each antigen has multiple epitopes, with one epitope shared across antigens \cite{Childs_Lauren_M2015-zn}. Epistasis in antigen-antibody binding interactions, quantified recently\cite{Adams2019-eb}, can also give rise to such disconnected sets.

Here, we take a phenomenological approach that is agnostic to molecular details. Exploiting Hopfield's \cite{Hopfield1982-fb} (or more generally, Gardner's \cite{Gardner1987-op}) construction, we construct fitness landscapes for each antigen with fitness islands around sequences corresponding to each epitope. In particular, consider $P$ epitopes on each antigen $\eta = 1$, $2$, that bind to antibody sequences $\mathbf{h}^{\eta}_{\alpha}$ ($\alpha = 1, \ldots P$). The fitness of an antibody with sequence $\mathbf{x}$ confronted by antigen $\eta$ is chosen to be $F^{\eta} \propto  \sum_{\alpha} ( \mathbf{x}\cdot \mathbf{h}^{\eta}_{\alpha} )^2$. This minimal construction produces fitness landscape with peaks at the specified epitopes $\mathbf{h}^{\eta}_{\alpha}$, provided $P$ is sufficiently small compared to sequence length $L$ \cite{Amit1985-as}.

By making different choices for the epitopes $\mathbf{h}^{\eta}_{\alpha}$, we may construct fitness landscapes with arbitrary amounts of correlation between them. We begin by studying the minimal case where $1$ epitope is shared between the two antigens, $\mathbf{h}^{1}_1 = \mathbf{h}^2_1$, with the others epitopes being uncorrelated. Later, we relax this assumption. For our theoretical analysis, we assume selection is strong and beneficial mutations are rapidly fixed, $s N \gg \mu N$, $ s N\gg 1$; hence fitness valleys between islands play a significant role.

\begin{figure*}
    \centering
    \includegraphics{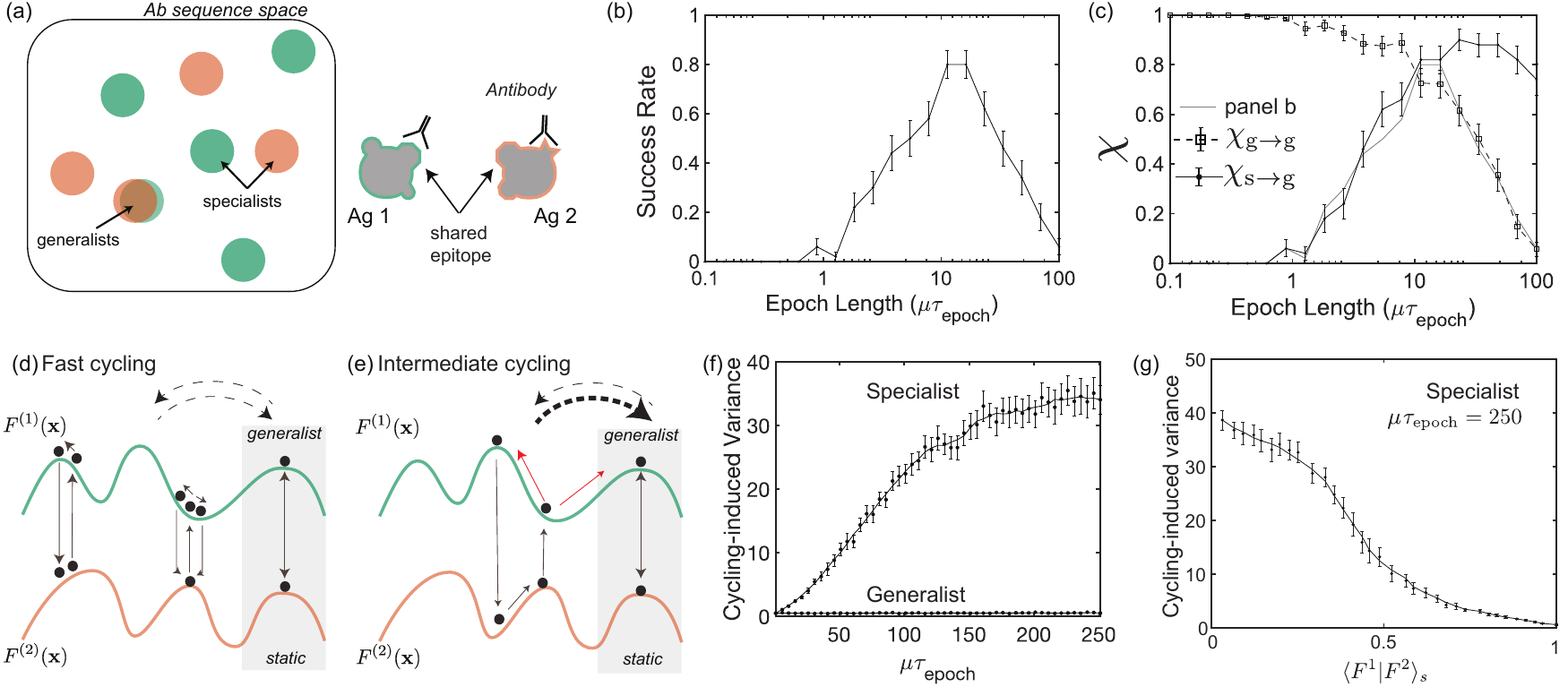}
    \caption{
    Intermediate timescale cycling enhances specialist-to-generalist conversions across fitness valleys without enhancing the time-reversed process. (a) Antibodies that bind distinct epitopes on antigens (bottom) form distinct specialist islands (top) in sequence space, separated by fitness valleys. Generalists bind an epitope shared by antigens. (b) Cycling at intermediate $\tau_{\text{epoch}}$ most reliably yields generalists in a finite population $N=100$ simulation. (c) Specialist-to-generalist transitions, $\chi_{\text{s} \to \text{g}}$, grows with $\tau_{\text{epoch}}$, while the ability to retain generalists $\chi_{\text{g} \to \text{g}}$ falls (both measured after $n=100$ cycles). (d) Fast cycling traps populations at fitness peaks near where they are initialized. (e) But intermediate $\tau_{\text{epoch}}$ allows evolution between specialists. Such evolution introduces sequence variance even in initially monoclonal specialist populations (red arrows in (e), quantified in (f)) but not for generalists.
    Such higher variance for specialists enhances specialists-to-generalists transitions but not the reverse process. (g) Cycling-induced variance is largest when specialists in $F^{(1)},F^{(2)}$ are uncorrelated (low $\langle F^{(1)} \vert F^{(2)} \rangle_s $).
    }
     \label{fig:Jijmodel}
\end{figure*}


We simulate a finite population of antibodies evolving via Moran dynamics. Initialising a monoclonal population at a specialist, we once again carried out simulations at different antigen switching times, $\tau_{\text{epoch}}$, and quantified the fraction of generalists in the population at long times. As seen in Fig.\ref{fig:Jijmodel}b, an intermediate timescale of switching elicits generalists in the population. This is reminiscent of the entropic model above, but for different underlying reasons.

Here, fast switching fails to produce generalists because populations stay confined to their initial position\cite{Weissman2009-uj} (Fig.\ref{fig:Jijmodel}d). In fact, as $\tau_{\text{epoch}} \to 0$, the fitness landscape is effectively static, $ F \sim F^{(1)} + F^{(2)}$, corresponding to the time-averaged fitness. This static landscape $F$ generally inherits the attractors of $F^{(1)}$ and $F^{(2)}$, as well as potentially numerous additional `spurious' attractors \cite{Amit1985-as}. Consequently, the population remains segregated away from the generalist genotypes by valleys of low-fitness, and generalist acquisition, $\chi_{\text{s} \to \text{g}}$, is small. In practice, such populations stuck in a specialist genotype for extended time can go extinct in the presence of multiple antigens \cite{Wang2015-jg}. 

In contrast, at slower switching times, evolution in each environment can shift the population away from its initial position in the prior environment (Fig.\ref{fig:Jijmodel}d). As shown in the SI, this requires at least time $\tau_{\text{min}} \sim d_{12}/\mu$, where $d_{12}$ is the typical mutational distance separating specialists across environments. Consequently, the population is forced to continually traverse genotype space. This continual evolution is by necessity stochastic (Fig.\ref{fig:Jijmodel}f), contingent on the random order of mutations that arise, as well as on any potential population variance. This cycling-induced mobility, augmented by stochasticity, allows the population to widely explore genotype space and find the generalist, and hence $\chi_{\text{s} \to \text{g}}$ rises (Fig.\ref{fig:Jijmodel}d).

Importantly, upon evolving into generalists, environmental cycling no longer disturbs the population, as the fitness of generalist sequences does not appreciably change over time. Thus cycling breaks the symmetry between specialists and generalists and enhances $\chi_{\text{s} \to \text{g}}$ without enhancing $\chi_{\text{g} \to \text{s}}$. Intuitively, intermediate cycling selectively `warms up' (i.e., increases stochasticity) specialist parts of sequence space, naturally leading the population to collect in `cooler' generalist sequences.


Cycling significantly slower than $\tau_{\text{min}}$ is counterproductive. The cycling-induced leaks from specialists to generalists only occur due to environmental switches; hence unnecessarily long $\tau_{\text{epoch}}$ only adds dead time with no additional population divergence. 

In the meantime, as shown in the SI, escape from generalists to specialists becomes significant on timescales of $(1/\mu) e^{\Delta F_g N}$ where $\Delta F_g$ is the fitness of the generalist relative to the fitness valley separating it from specialists; $N$ is the population size. See \cite{Jain2007-tl,Van_Nimwegen2000-qh,Weissman2009-uj} for calculations of valley crossing rates in other parameter regimes. These considerations limit intermediate timescales favorable for evolving generalists:


\begin{equation}
\tau_{\text{min}} \sim  d_{12}/ \mu \quad  < \quad  \tau_{\text{epoch}}  \quad <\quad  \tau_{\text{max}} \sim (1/\mu) e^{\Delta F_g N}
\end{equation}

\subsubsection*{Correlation between specialists}

The effectiveness of this theoretical cycling mechanism depends on the correlation between specialists of $F^{(1)}$ and $F^{(2)}$ \cite{Wang2019-al}. For example, if specialists of $F^{(1)}$ and $F^{(2)}$ are similar or are carefully arranged as to be well within each other’s attractors, cycling will primarily cycle the population between specialists with minimal divergence into generalists. As shown in the SI, we can quantify relevant correlations by 
$$ \langle F^{(1)} \vert F^{(2)} \rangle_s \equiv \f{c_{1,2}}{\sqrt{c_{1,1} \, c_{2,2}}} $$ 
where $ c_{\eta, \gamma} = \f{1}{L P}\sum_{\alpha,\beta \neq 1} \mathbf{h}^{\eta}_{\alpha}\cdot \mathbf{h}^{\gamma}_{\beta}$ excludes the generalist pattern $\mathbf{h}^{1}_1 = \mathbf{h}^{2}_1$. When $\langle F^{(1)} \vert F^{(2)} \rangle_s$ is high, cycling-induced variance is low; see Fig.\ref{fig:Jijmodel}g.  Consequently, the small asymmetry between $\chi_{\text{s} \to \text{g}}$ and $\chi_{\text{g} \to \text{s}}$ created by a single environmental cycle must be compounded by cycling multiple times; however, in practice, other considerations might limit the number of such cycles. Hence, our proposal requires the specialists of $F^{(1)}$ and $F^{(2)}$ to be sufficiently uncorrelated (low $\langle F^{(1)} \vert F^{(2)} \rangle_s$).

\begin{figure}
\begin{centering}
\includegraphics[width=\columnwidth]{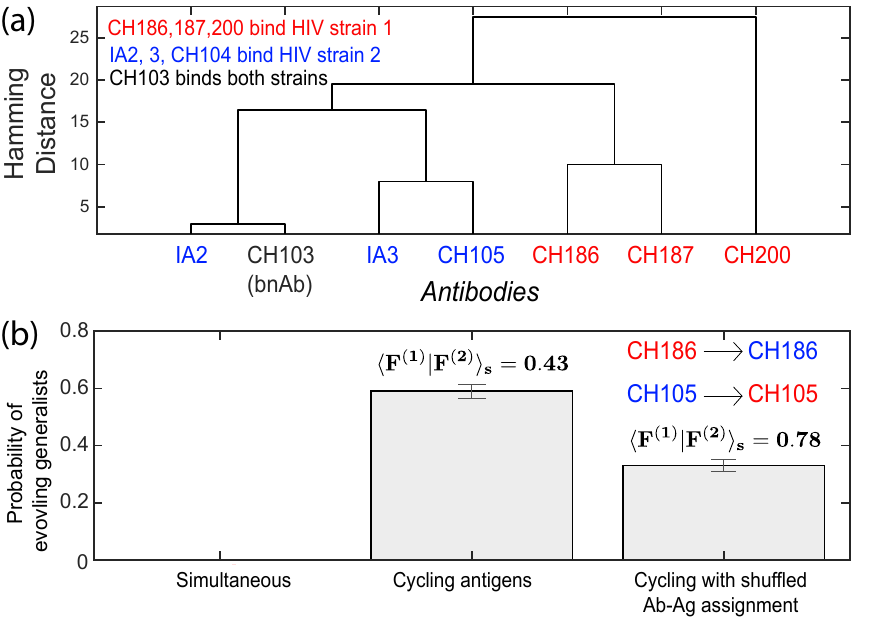}
\caption{Cycling between fitness landscapes constructed using antibody sequences from HIV patients yields generalists; however, cycling is less effective for artificially shuffled data with higher specialist correlation. (a) Sequence divergence of antibodies that bind two distinct strains (red, blue) of HIV. See SI for sequence and binding affinity data, reproduced from \cite{Liao2013-xu,Gao2014-uy,Bonsignori2016-du}. (b) Following Gardner\cite{Gardner1987-op}, we constructed two fitness landscapes $F^{(1)},F^{(2)}$ with peaks at red, blue sequences resp. and simulated evolution with realistic parameters (see SI). Generalists are evolved only if antigens are cycled.  Cycling is less effective if we shuffle antibody-antigen assignment: CH105 now considered specialized for strain 2 (i.e., now red), CH186 for strain 1 (i.e., now blue). Shuffling artificially increases specialist correlation $\langle F^{(1)} \vert F^{(2)} \rangle_s$ from $0.43$ to $0.78$. 
\label{fig:data}}
\end{centering}
\end{figure}

Is cycling a practical strategy given physiological parameters for population dynamics and the correlations between specialists antibodies found during HIV infection?
We analyzed specialist and generalist antibody sequences collected from an HIV patient \cite{Liao2013-xu, Gao2014-uy, Bonsignori2016-du}; see Fig.\ref{fig:data}a. We constructed landscapes $F^{(1)},F^{(2)}$ with fitness peaks at these observed specialist and generalist sequences following Gardner's construction \cite{Gardner1987-op}; see SI for details.

Simulations of cycling environments $F^{(1)},F^{(2)}$ constructed from the above sequence data evolved generalist antibodies, while simultaneous presentation of both antigens, a practical alternative to fast cycling\cite{Wang2015-jg}, fails to produce such generalists; see Fig.\ref{fig:data}b. We then artificially shuffled antigen labels for antibodies, so that CH105 was considered a Ag2 specialist and CH186, an Ag1 specialist and reconstructed $F^{(1)},F^{(2)}$. This artificial shuffling significantly increased the correlation $\langle F^{(1)} \vert F^{(2)} \rangle_s = 0.78$ compared to the real data ($\langle F^{(1)} \vert F^{(2)} \rangle_s = 0.43$). Cycling is now less effective in evolving generalists. We conclude that the low correlation between specialists in the real data is crucial for time-varying selection of generalists.

While our model here did not explicitly account for extinction, simultaneous presentation or fast cycling can cause most specialist B-cells to perish, especially if many distinct antigens are use (see SI). In this more realistic case, `chirped' cycling at increasing frequency as in Fig.\ref{fig:chirp} will provide the dual advantage of evolving generalists from specialists through slow cycling and then enhance generalist yield by removing specialists through fast cycling.

\section*{Discussion}

%

We have shown that environmental changes on intermediate timescales can dynamically funnel populations from specialists to generalists. Our  quantitative framework suggests broad new classes of experimental protocols such as `chirped cycling' that further enhance the evolution of generalists. 

The relevant intermediate timescale here is that of evolutionary transients in a population - the environment must change slow enough for significant changes to accumulate but fast enough to prevent the population from settling to a steady state distribution.
This intermediate regime creates a highly dynamic fitness `seascape' with no effective static description\cite{Mustonen2009-fu}. This dynamic regime has been relatively less explored\cite{Cvijovic2015-xy} than limits where the environment changes much faster or slower than evolutionary transients and which can be understood using effective static environments.  

The simple models studied here ignore many ingredients present in B-cell affinity maturation and other evolutionary processes in the natural world. For instance, affinity maturation starts from a specific naive antibody repertoire\cite{Elhanati_Yuval2015-xh}, population response timescales can vary widely \cite{Hallatschek2018-dq}; our results require an ensemble of lineages to participate\cite{Wang2015-jg} and ignores clonal interference \cite{Park2007-ke}.

Nonetheless, our analysis has broad applicability, since it relies only on a simple phenomenological characterization of how specialist and generalist genotypes are organized in sequence space.  Further, our results are fundamentally linked to the fact that generalists experience less time variation of fitness than specialists, leading, e.g., to higher stochasticity and mobility for the specialist parts of sequence space but not for the generalists. In this sense, the dynamic strategies presented here represent a broader class of non-equilibrium evolutionary strategies\cite{Mustonen2009-fu,Kussell2014-mg} that can enhance the rate of transitions from specialists to generalists without enhancing the time-reversed processes. 

Dynamic protocols have been investigated recently in other evolutionary contexts, e.g., in antibiotic resistance, where correlations in response to different antibiotics have been exploited to maximize cross-vulnerability\cite{Marrec2018-vv,Munck2014-vr, Chait2007-bb,Nichol2015-zg}. While such cross-vulnerabilities have been primarily studied in the slow switching limit, switching antibiotics after a partial evolutionary response like that explored here might open a larger space of strategies. 

While we have discussed dynamic environments as a prescriptive mechanism, natural environments are also dynamic \cite{Levins1968-qz}.
For example, co-evolution of pathogens \cite{Papkou2019-ur, Nourmohammad2016-jm}, movement through spatially heterogeneous environments \cite{Zhang2011-wa} and ecological changes \cite{Pelletier_F2009-zg,Roxburgh2004-wl}) can naturally create the intermediate timescale variations discussed here. The quantitative principles developed here suggest new experiments to both understand and exploit this understudied regime of evolution with no separation of timescales between perturbation and response.


\section*{Acknowledgements}
We thank Sarah Cobey, Aaron Dinner, Allan Drummond,  Muhittin Mungan, Sidney Nagel, Stephanie Palmer, David Pincus, Rama Ranganathan, Olivier Rivoire, Thomas Witten for useful discussions. VS thanks the NIH for support through NIBIB T32 EB009412. KBH and AM thank the James S. McDonnell Foundation and the Simons Foundation respectively for support. SW is grateful to funding from UCLA.

\bibliography{Paperpile}

\begin{thebibliography}{10}

\bibitem{Levins1968-qz}
Richard Levins.
\newblock {\em Evolution in Changing Environments: Some Theoretical
  Explorations}.
\newblock Princeton University Press, August 1968.

\bibitem{Burton2004-jw}
Dennis~R Burton, Ronald~C Desrosiers, Robert~W Doms, et~al.
\newblock {HIV} vaccine design and the neutralizing antibody problem.
\newblock {\em Nat. Immunol.}, 5(3):233--236, March 2004.

\bibitem{Burton2012-zd}
Dennis~R Burton, Pascal Poignard, Robyn~L Stanfield, and Ian~A Wilson.
\newblock Broadly neutralizing antibodies present new prospects to counter
  highly antigenically diverse viruses.
\newblock {\em Science}, 337(6091):183--186, July 2012.

\bibitem{Wu2010-xh}
Xueling Wu, Zhi-Yong Yang, Yuxing Li, et~al.
\newblock Rational design of envelope identifies broadly neutralizing human
  monoclonal antibodies to {HIV-1}.
\newblock {\em Science}, 329(5993):856--861, August 2010.

\bibitem{Cobey_Sarah2015-tb}
{Cobey Sarah}, {Wilson Patrick}, and {Matsen Frederick A.}
\newblock The evolution within us.
\newblock {\em Philos. Trans. R. Soc. Lond. B Biol. Sci.}, 370(1676):20140235,
  September 2015.

\bibitem{Pissani2012-to}
Franco Pissani, Delphine~C Malherbe, Harlan Robins, et~al.
\newblock Motif-optimized subtype a {HIV} envelope-based {DNA} vaccines rapidly
  elicit neutralizing antibodies when delivered sequentially.
\newblock {\em Vaccine}, 30(37):5519--5526, August 2012.

\bibitem{Malherbe2011-cg}
Delphine~C Malherbe, Nicole~A Doria-Rose, Lynda Misher, et~al.
\newblock Sequential immunization with a subtype {B} {HIV-1} envelope
  quasispecies partially mimics the in vivo development of neutralizing
  antibodies.
\newblock {\em J. Virol.}, 85(11):5262--5274, June 2011.

\bibitem{Wang2017-tx}
Shenshen Wang.
\newblock Optimal sequential immunization can focus antibody responses against
  diversity loss and distraction.
\newblock {\em PLoS Comput. Biol.}, 13(1):e1005336, January 2017.

\bibitem{Wang2015-jg}
Shenshen Wang, Jordi Mata-Fink, Barry Kriegsman, et~al.
\newblock Manipulating the selection forces during affinity maturation to
  generate cross-reactive {HIV} antibodies.
\newblock {\em Cell}, 160(4):785--797, February 2015.

\bibitem{Mustonen2010-kf}
Ville Mustonen and Michael L{\"a}ssig.
\newblock Fitness flux and ubiquity of adaptive evolution.
\newblock {\em Proc. Natl. Acad. Sci. U. S. A.}, 107(9):4248--4253, March 2010.

\bibitem{Mustonen2009-fu}
Ville Mustonen and Michael L{\"a}ssig.
\newblock From fitness landscapes to seascapes: non-equilibrium dynamics of
  selection and adaptation.
\newblock {\em Trends Genet.}, 25(3):111--119, March 2009.

\bibitem{Arndt2004-yq}
Peter~F Arndt and Terence Hwa.
\newblock Regional and time-resolved mutation patterns of the human genome.
\newblock {\em Bioinformatics}, 20(10):1482--1485, July 2004.

\bibitem{Kussell2014-mg}
E~Kussell and M~Vucelja.
\newblock Non-equilibrium physics and evolution--adaptation, extinction, and
  ecology: a key issues review.
\newblock {\em Rep. Prog. Phys.}, 77(10):102602, October 2014.

\bibitem{Goldenfeld2011-pm}
Nigel Goldenfeld and Carl Woese.
\newblock Life is physics: Evolution as a collective phenomenon far from
  equilibrium.
\newblock {\em Annu. Rev. Condens. Matter Phys.}, February 2011.

\bibitem{Cvijovic2015-xy}
Ivana Cvijovi{\'c}, Benjamin~H Good, Elizabeth~R Jerison, and Michael~M Desai.
\newblock Fate of a mutation in a fluctuating environment.
\newblock {\em Proc. Natl. Acad. Sci. U. S. A.}, 112(36):E5021--8, September
  2015.

\bibitem{Toprak2011-ae}
Erdal Toprak, Adrian Veres, Jean-Baptiste Michel, et~al.
\newblock Evolutionary paths to antibiotic resistance under dynamically
  sustained drug selection.
\newblock {\em Nat. Genet.}, 44(1):101--105, December 2011.

\bibitem{Marrec2018-vv}
Lo{\"\i}c Marrec and Anne-Florence Bitbol.
\newblock Quantifying the impact of a periodic presence of antimicrobial on
  resistance evolution in a homogeneous microbial population of fixed size.
\newblock {\em J. Theor. Biol.}, 457:190--198, November 2018.

\bibitem{De_Jong2018-ao}
Maxwell~G De~Jong and Kevin~B Wood.
\newblock Tuning spatial profiles of selection pressure to modulate the
  evolution of drug resistance.
\newblock {\em Phys. Rev. Lett.}, 120(23):238102, June 2018.

\bibitem{Gatenby2009-hp}
Robert~A Gatenby, Ariosto~S Silva, Robert~J Gillies, and B~Roy Frieden.
\newblock Adaptive therapy.
\newblock {\em Cancer Res.}, 69(11):4894--4903, June 2009.

\bibitem{Katouli2011-uk}
Allen~A Katouli and Natalia~L Komarova.
\newblock The worst drug rule revisited: mathematical modeling of cyclic cancer
  treatments.
\newblock {\em Bull. Math. Biol.}, 73(3):549--584, March 2011.

\bibitem{Kassen2002-co}
R~Kassen.
\newblock The experimental evolution of specialists, generalists, and the
  maintenance of diversity: Experimental evolution in variable environments.
\newblock {\em J. Evol. Biol.}, 15(2):173--190, March 2002.

\bibitem{Desponds2016-ko}
Jonathan Desponds, Thierry Mora, and Aleksandra~M Walczak.
\newblock Fluctuating fitness shapes the clone-size distribution of immune
  repertoires.
\newblock {\em Proc. Natl. Acad. Sci. U. S. A.}, 113(2):274--279, January 2016.

\bibitem{Uecker2011-kn}
Hildegard Uecker and Joachim Hermisson.
\newblock On the fixation process of a beneficial mutation in a variable
  environment.
\newblock {\em Genetics}, 188(4):915--930, August 2011.

\bibitem{Hemery2015-hv}
Mathieu Hemery and Olivier Rivoire.
\newblock Evolution of sparsity and modularity in a model of protein allostery.
\newblock {\em Phys. Rev. E Stat. Nonlin. Soft Matter Phys.}, 91(4):042704,
  April 2015.

\bibitem{Kashtan2005-zp}
Nadav Kashtan and Uri Alon.
\newblock Spontaneous evolution of modularity and network motifs.
\newblock {\em Proc. Natl. Acad. Sci. U. S. A.}, 102(39):13773--13778,
  September 2005.

\bibitem{Lipson2002-un}
Hod Lipson, Jordan~B Pollack, and Nam~P Suh.
\newblock On the origin of modular variation.
\newblock {\em Evolution}, 56(8):1549--1556, August 2002.

\bibitem{Xue2016-al}
Bingkan Xue and Stanislas Leibler.
\newblock Evolutionary learning of adaptation to varying environments through a
  transgenerational feedback.
\newblock {\em Proceedings of the National Academy of Sciences},
  113(40):11266--11271, October 2016.

\bibitem{Raman2016-bj}
Arjun~S Raman, K~Ian White, and Rama Ranganathan.
\newblock Origins of allostery and evolvability in proteins: A case study.
\newblock {\em Cell}, 166(2):468--480, July 2016.

\bibitem{Mustonen2008-im}
Ville Mustonen and Michael L{\"a}ssig.
\newblock Molecular evolution under fitness fluctuations.
\newblock {\em Phys. Rev. Lett.}, 100(10):108101, March 2008.

\bibitem{Kussell2006-ch}
Edo Kussell, Stanislas Leibler, and Alexander Grosberg.
\newblock Polymer-population mapping and localization in the space of
  phenotypes.
\newblock {\em Phys. Rev. Lett.}, 97(6):068101, August 2006.

\bibitem{Mayer2017-gl}
Andreas Mayer, Thierry Mora, Olivier Rivoire, and Aleksandra~M Walczak.
\newblock Transitions in optimal adaptive strategies for populations in
  fluctuating environments.
\newblock {\em Phys Rev E}, 96(3-1):032412, September 2017.

\bibitem{Munoz2005-am}
Enrique~T Mu{\~n}oz and Michael~W Deem.
\newblock Epitope analysis for influenza vaccine design.
\newblock {\em Vaccine}, 23(9):1144--1148, January 2005.

\bibitem{Chaudhury2014-tq}
Sidhartha Chaudhury, Jaques Reifman, and Anders Wallqvist.
\newblock Simulation of {B} cell affinity maturation explains enhanced antibody
  cross-reactivity induced by the polyvalent malaria vaccine {AMA1}.
\newblock {\em J. Immunol.}, 193(5):2073--2086, September 2014.

\bibitem{Deem2003-yu}
Michael~W Deem and Ha~Youn Lee.
\newblock Sequence space localization in the immune system response to
  vaccination and disease.
\newblock {\em Phys. Rev. Lett.}, 91(6):068101, August 2003.

\bibitem{Childs_Lauren_M2015-zn}
{Childs Lauren M.}, {Baskerville Edward B.}, and {Cobey Sarah}.
\newblock Trade-offs in antibody repertoires to complex antigens.
\newblock {\em Philos. Trans. R. Soc. Lond. B Biol. Sci.}, 370(1676):20140245,
  September 2015.

\bibitem{Pressman2019-at}
Abe~D Pressman, Ziwei Liu, Evan Janzen, et~al.
\newblock Mapping a systematic ribozyme fitness landscape reveals a frustrated
  evolutionary network for {Self-Aminoacylating} {RNA}.
\newblock {\em J. Am. Chem. Soc.}, 141(15):6213--6223, April 2019.

\bibitem{Blanco2019-to}
Celia Blanco, Evan Janzen, Abe Pressman, Ranajay Saha, and Irene~A Chen.
\newblock Molecular fitness landscapes from {High-Coverage} sequence profiling.
\newblock {\em Annu. Rev. Biophys.}, 48:1--18, May 2019.

\bibitem{Adams2019-eb}
Rhys~M Adams, Justin~B Kinney, Aleksandra~M Walczak, and Thierry Mora.
\newblock Epistasis in a fitness landscape defined by {Antibody-Antigen}
  binding free energy.
\newblock {\em Cell Syst}, 8(1):86--93.e3, January 2019.

\bibitem{Hopfield1982-fb}
J~J Hopfield.
\newblock Neural networks and physical systems with emergent collective
  computational abilities.
\newblock In {\em Proceedings of the International Association for Shell and
  Spatial Structures ({IASS}) Symposium 2009}, January 1982.

\bibitem{Gardner1987-op}
Elizabeth Gardner.
\newblock Maximum storage capacity in neural networks.
\newblock {\em EPL}, 4(4):481, 1987.

\bibitem{Amit1985-as}
Daniel Amit, Hanoch Gutfreund, and H~Sompolinsky.
\newblock Storing infinite numbers of patterns in a spin-glass model of neural
  networks.
\newblock {\em Phys. Rev. Lett.}, 55(14):1530--1533, September 1985.

\bibitem{Weissman2009-uj}
Daniel~B Weissman, Michael~M Desai, Daniel~S Fisher, and Marcus~W Feldman.
\newblock The rate at which asexual populations cross fitness valleys.
\newblock {\em Theor. Popul. Biol.}, 75(4):286--300, June 2009.

\bibitem{Jain2007-tl}
Kavita Jain and Joachim Krug.
\newblock Deterministic and stochastic regimes of asexual evolution on rugged
  fitness landscapes.
\newblock {\em Genetics}, 175(3):1275--1288, March 2007.

\bibitem{Van_Nimwegen2000-qh}
E~van Nimwegen and J~P Crutchfield.
\newblock Metastable evolutionary dynamics: crossing fitness barriers or
  escaping via neutral paths?
\newblock {\em Bull. Math. Biol.}, 62(5):799--848, September 2000.

\bibitem{Wang2019-al}
Shenshen Wang and Lei Dai.
\newblock Evolving generalists via dynamic sculpting of rugged landscapes.
\newblock May 2019.

\bibitem{Liao2013-xu}
Hua-Xin Liao, Rebecca Lynch, Tongqing Zhou, et~al.
\newblock Co-evolution of a broadly neutralizing {HIV-1} antibody and founder
  virus.
\newblock {\em Nature}, 496(7446):469--476, April 2013.

\bibitem{Gao2014-uy}
Feng Gao, Mattia Bonsignori, Hua-Xin Liao, et~al.
\newblock Cooperation of {B} cell lineages in induction of {HIV-1-broadly}
  neutralizing antibodies.
\newblock {\em Cell}, 158(3):481--491, July 2014.

\bibitem{Bonsignori2016-du}
Mattia Bonsignori, Tongqing Zhou, Zizhang Sheng, et~al.
\newblock Maturation pathway from germline to broad {HIV-1} neutralizer of a
  {CD4-Mimic} antibody.
\newblock {\em Cell}, 165(2):449--463, April 2016.

\bibitem{Elhanati_Yuval2015-xh}
{Elhanati Yuval}, {Sethna Zachary}, {Marcou Quentin}, et~al.
\newblock Inferring processes underlying b-cell repertoire diversity.
\newblock {\em Philos. Trans. R. Soc. Lond. B Biol. Sci.}, 370(1676):20140243,
  September 2015.

\bibitem{Hallatschek2018-dq}
Oskar Hallatschek.
\newblock {Selection-Like} biases emerge in population models with recurrent
  jackpot events.
\newblock {\em Genetics}, 210(3):1053--1073, November 2018.

\bibitem{Park2007-ke}
Su-Chan Park and Joachim Krug.
\newblock Clonal interference in large populations.
\newblock {\em Proc. Natl. Acad. Sci. U. S. A.}, 104(46):18135--18140, November
  2007.

\bibitem{Munck2014-vr}
Christian Munck, Heidi~K Gumpert, Annika I~Nilsson Wallin, Harris~H Wang, and
  Morten O~A Sommer.
\newblock Prediction of resistance development against drug combinations by
  collateral responses to component drugs.
\newblock {\em Sci. Transl. Med.}, 6(262):262ra156, November 2014.

\bibitem{Chait2007-bb}
Remy Chait, Allison Craney, and Roy Kishony.
\newblock Antibiotic interactions that select against resistance.
\newblock {\em Nature}, 446(7136):668--671, April 2007.

\bibitem{Nichol2015-zg}
Daniel Nichol, Peter Jeavons, Alexander~G Fletcher, et~al.
\newblock Steering evolution with sequential therapy to prevent the emergence
  of bacterial antibiotic resistance.
\newblock {\em PLoS Comput. Biol.}, 11(9):e1004493, September 2015.

\bibitem{Papkou2019-ur}
Andrei Papkou, Thiago Guzella, Wentao Yang, et~al.
\newblock The genomic basis of red queen dynamics during rapid reciprocal
  host-pathogen coevolution.
\newblock {\em Proc. Natl. Acad. Sci. U. S. A.}, 116(3):923--928, January 2019.

\bibitem{Nourmohammad2016-jm}
Armita Nourmohammad, Jakub Otwinowski, and Joshua~B Plotkin.
\newblock {Host-Pathogen} coevolution and the emergence of broadly neutralizing
  antibodies in chronic infections.
\newblock {\em PLoS Genet.}, 12(7):e1006171, July 2016.

\bibitem{Zhang2011-wa}
Qiucen Zhang, Guillaume Lambert, David Liao, et~al.
\newblock Acceleration of emergence of bacterial antibiotic resistance in
  connected microenvironments.
\newblock {\em Science}, 333(6050):1764--1767, September 2011.

\bibitem{Pelletier_F2009-zg}
{Pelletier F.}, {Garant D.}, and {Hendry A.P.}
\newblock Eco-evolutionary dynamics.
\newblock {\em Philos. Trans. R. Soc. Lond. B Biol. Sci.},
  364(1523):1483--1489, June 2009.

\bibitem{Roxburgh2004-wl}
Stephen~H Roxburgh, Katriona Shea, and J~Bastow Wilson.
\newblock {THE} {INTERMEDIATE} {DISTURBANCE} {HYPOTHESIS}: {PATCH} {DYNAMICS}
  {AND} {MECHANISMS} {OF} {SPECIES} {COEXISTENCE}.
\newblock {\em Ecology}, 85(2):359--371, February 2004.

\end{thebibliography}


\begin{thebibliography}{10}

\bibitem{Wang2015}
Shenshen Wang, Jordi Mata-Fink, Barry Kriegsman, et~al.
\newblock Manipulating the selection forces during affinity maturation to
  generate cross-reactive hiv antibodies.
\newblock {\em Cell}, 160(4):785 -- 797, 2015.

\bibitem{Maruvka2011}
Yosef~E. Maruvka, David~A. Kessler, and Nadav~M. Shnerb.
\newblock The birth-death-mutation process: A new paradigm for fat tailed
  distributions.
\newblock {\em PLOS ONE}, 6(11):1--7, 11 2011.

\bibitem{moran1958}
P.~A.~P. Moran.
\newblock Random processes in genetics.
\newblock {\em Mathematical Proceedings of the Cambridge Philosophical
  Society}, 54(1):60–71, 1958.

\bibitem{Weissman2009}
Daniel~B Weissman, Michael~M Desai, Daniel~S Fisher, and Marcus~W Feldman.
\newblock The rate at which asexual populations cross fitness valleys.
\newblock {\em Theor. Popul. Biol.}, 75(4):286--300, June 2009.

\bibitem{Van_Nimwegen2000}
E~van Nimwegen and J~P Crutchfield.
\newblock Metastable evolutionary dynamics: crossing fitness barriers or
  escaping via neutral paths?
\newblock {\em Bull. Math. Biol.}, 62(5):799--848, September 2000.

\bibitem{Jain2007}
Kavita Jain and Joachim Krug.
\newblock Deterministic and stochastic regimes of asexual evolution on rugged
  fitness landscapes.
\newblock {\em Genetics}, 175(3):1275--1288, March 2007.

\bibitem{Metropolis1953}
Nicholas Metropolis, Arianna~W. Rosenbluth, Marshall~N. Rosenbluth, Augusta~H.
  Teller, and Edward Teller.
\newblock Equation of state calculations by fast computing machines.
\newblock {\em The Journal of Chemical Physics}, 21(6):1087--1092, 1953.

\bibitem{Bonsignori2016}
Mattia Bonsignori, Tongqing Zhou, Zizhang Sheng, et~al.
\newblock Maturation pathway from germline to broad hiv-1 neutralizer of a
  cd4-mimic antibody.
\newblock {\em Cell}, 165(2):449 -- 463, 2016.

\bibitem{Gao2014}
Feng Gao, Mattia Bonsignori, Hua-Xin Liao, et~al.
\newblock Cooperation of b cell lineages in induction of hiv-1-broadly
  neutralizing antibodies.
\newblock {\em Cell}, 158(3):481 -- 491, 2014.

\bibitem{Liao2013}
Hua-Xin Liao, Rebecca Lynch, Tongqing Zhou, et~al.
\newblock Co-evolution of a broadly neutralizing hiv-1 antibody and founder
  virus.
\newblock {\em Nature}, 496:469 EP --, 04 2013.

\end{thebibliography}
\bibliographystyle{test}
\end{document}


\title{ {\Large  Tuning environmental timescales to evolve and maintain generalists} \\ {\normalsize SI} }

\author{Vedant Sachdeva, Kabir Husain, Jiming Sheng, Shenshen Wang, Arvind Murugan}

\maketitle 

\section{Entropically Disfavored Generalists}

\subsection{Model}


To construct landscapes with entropically disfavored generalists, we model antibodies and antigens in a manner similar to the one described by by Wang et. al\cite{Wang2015}. In this model, the sequence of each antibody, $\mathbf{x}$ is a sequence of length $L$ with entries $\pm1$. Each antigen, indexed by $\eta$, is assumed to have an epitope of sequence, $\mathbf{h}^{\eta}$. Each epitope is length $L$ with entries $\pm1$. The binding affinity of a given antigen to an antibody is given by an additive sum-over-sites model:

\be \label{eq:hisienergy}
E \l \mathbf{x}, \mathbf{h^{\eta}} \r =   \sum_{i}^{L} {h}^{\eta}_{i} {x}_{i}  
\ee

The fitness of each antibody $\mathbf{x}$ in the presence with antigen $\eta$ is given by thresholding its binding affinity, as follows:

\be \label{eq:hisiLandscape}
F^{(\eta)} \l \mathbf{x} \r = s \epsilon \Theta \l E(\mathbf{x},\mathbf{h}^{\eta}) - T \r - s
\ee

Here, $\Theta (x)$ is the Heaviside function. $T$ is the binding affinity threshold that an antibody must overcome before reaping a fitness benefit. By construction, all fit individuals have the same fitness $s (\epsilon-1) > 0$ and all unfit individuals are equally unfit to an extent $-s$, resulting in a degeneracy of fit and unfit genotypes.


\subsection{Specialists and Generalists}

Biological constraints impose that parts of an antigen's epitope is conserved, while other parts are variable as the viral strain evolves. As such, we suppose $L_{c}$ sites of the $L$ total sites are fixed across $\eta$, while the others are unique to each antigen. This constraint results in the possibility that some antibody sequences have a positive fitness for all antigens. Such antibodies are called generalists. The number of generalists  is a function of the threshold $T$ and the lengths of the conserved $L_c$ and variable $L_{v}=L-L_c$ regions. In particular, if $(T+L)/2>L_{c}+\frac{1}{2}L_{v}$, there are no generalists. A simple equation to compute the number of generalists against the number of viable antibodies is given as follows:

\begin{gather}
    \text{Fraction of Generalists}=\frac{\sum_{j=0}^{L} {L_{c}\choose{j}}{L_{v}\choose{(T+L)/2-j}}}{\sum_{k=(T+L)/2}^{L} {L\choose{k}}}
\end{gather}

\noindent where, ${{N}\choose{m}}$ is the combinatorial function. By rule, this function is zero if $m>N$ or $m<0$. 

Here, we considered $L=19$, $L_{c}=12$, and $T=11$. The proportion of viable antibodies that are generalists for these parameter choices is $\approx1.3\%$. This choice is qualitatively similar to the analysis developed in \cite{Wang2015} based on experiments there.

\subsection{Finite population simulation}\label{sec:finitepopsim}

We simulate a population of antibodies evolving in the landscapes Eq. \ref{eq:hisiLandscape}, subject to mutation, growth, and death. In particular, we use a `Yule' process \cite{Maruvka2011} with the following stochastic events:

\begin{itemize}
	\item \textbf{Mutation} with rate $\mu$ per individual, in which a single site on the genome is mutated (i.e. a single bit-flip of $\mathbf{x}$).
	\item \textbf{Birth-death} with rate $\lambda$ per individual, in which the individual is duplicated with probability:
	\begin{gather}
	\text{Pr} \l \mathbf{x} \text{ reproduces} \r =\Theta \l E(\mathbf{x}, h^{\eta}) - T \r \l 1 - \f{N}{K} \r 
	\end{gather}
	If the individual does not reproduce, it is removed from the population. Here, $N$ is the current population size, $K=500$ is a carrying capacity that prevents the population from growing indefinitely, and $F^{(\eta)}$ is the fitness of that individual in the (current) landscape $\eta$.
\end{itemize}

\indent At each event, time is advanced by the usual exponentially distributed amount. The environment $\eta$ is taken to alternate between $\eta = 1$ and $\eta = 2$ every $\tau_{\text{epoch}}$.

\indent Choosing units of time by setting the birth-death rate $\lambda = 1$, we set the mutation rate to $\mu = 0.05$. Finally, we set the carrying capacity $K = 500$. We can infer from these choices that $s = 1, \epsilon = 2-\frac{2N}{K}$.

We evolved an initially monoclonal population of size $N = 10$ (initialised with $\mathbf{x} = \mathbf{h}^{\eta}$ for all individuals). Simulations were run for either a fixed time $t = 100$ (in units of $\lambda$) or for $t = 10 \tau_{\text{epoch}}$, whichever is longer, and we performed $25$ replicates for each value of $\tau_{\text{epoch}}$. For each run, we saved the number of generalists and the overall population size at the end of the simulation. In Fig. 2b of the main text, we plotted the proportion of trials that had more than $10$ generalist antibodies at the end of the simulation, finding that the proportion was high for an intermediate rate of cycling. 

\subsection{Evolution between Specialists and Generalists: $\chi_{\text{s}\to\text{g}}$ and $\chi_{\text{g}\to\text{g}}$}

As discussed in the main text, there are two possible failure modes: (1) at cycling rates too fast, the population does not have the time to evolve a generalist before adverse selection result in population extinction, (2) at cycling rates too slow, the population loses its ability to maintain generalists. In order to illustrate the tension between cycling too fast and cycling too slow, we compute two quantities:
\begin{itemize}
    \item $\chi_{\text{g}\to\text{g}}$, the fraction of generalists remaining a generalist after one epoch. We define this quantity as the fraction of an initially monoclonal generalist population remaining a generalist after a single epoch.
    \item $\chi_{\text{s}\to\text{g}}$, the fraction of trials in which a monoclonal population of specialists, initialized at $\mathbf{x}=\mathbf{h}^{(1)}$, evolve a single generalist within an epoch. 
\end{itemize}

\indent Both $\chi_{\text{g}\to\text{g}}$, $\chi_{\text{s}\to\text{g}}$ are computed from $50$ replicates for each $\tau_{\text{epoch}}$. As plotted in Fig. 2c, we observe that $\chi_{\text{s}\to\text{g}}$ starts initially at $0$ and rises with $\tau_{\text{epoch}}$, while $\chi_{\text{g}\to\text{g}}$ starts at $1$ and falls with $\tau_{\text{epoch}}$.\\

\textbf{Population traces:} Fig. 2d shows population traces in single runs. As in Section \ref{sec:finitepopsim}, we initialized a population at $\mathbf{x}=\mathbf{h}^{(1)}$ with parameters as above. Generalist fraction is defined as $\frac{\text{number of generalists}}{\text{population size}}$.

We use the following values of $\tau_{\text{epoch}}$ for Fig.2(d):
 Fast cycling $\tau_{\text{epoch}}=1$ (Fig. 2d(i)), Intermediate cycling $\tau_{\text{epoch}}=60$ (Fig.2d(ii)), Slow cycling  $\tau_{\text{epoch}}=400$ (Fig.2d(iii))

\subsection{Timescale Analysis}

Our numerical study found that an intermediate timescale of environmental cycling, $\tau_{\text{min}} < \tau_{\text{epoch}} < \tau_{\text{max}}$, was most effective at obtaining generalists. Here we estimate the bounds, $\tau_{\text{min}}$and $\tau_{\text{max}}$, in terms of the mutation rate $\mu$, the population size $N$, the length of the genotype $L$, and the distance from the initial ancestral genotype to the generalist, $d_{i \to g}$.

\subsubsection*{Finding the generalist: $\tau_{\text{min}}$}

Consider a population initialised as a specialist for antigen $1$. For sufficiently strong selection pressure ($s > \mu \log N$), purifying selection drives the population to extinction if a generalist has not been discovered before the environment switches to antigen $2$. Thus we demand that $\tau_{\text{epoch}}$ is long enough for the population to evolve a generalist in a single epoch.

As fitness is uniform across the specialist region, the population must discover the generalist by diffusion. An initially monoclonal population of size $N$ diffuses out from the initial genotype. If the population size is much smaller than the number of possible genotypes ($N \ll 2^L$), there are two possible regimes of the diffusive search:

\begin{enumerate}
	\item The initial genotype is far from the generalist: more precisely, the population size $N$ is smaller than the set of sequences between the initial genotype and the generalist. In terms of the Hamming distance between the initial genotype to the generalist, $d_{i \to g}$:

	\be
	\sum_{d = 0}^{d_{i \to g}} \binom{L}{d} > N
	\ee

	\indent In this regime, finding the generalist is a rare event, requiring time $\mu \tau_{\text{min}} \sim 2^L$, i.e. the time taken to explore all of genotype space. It is therefore extremely improbable that the generalist will be found, and population extinction is likely.

	\item The initial genotype is close to the generalist: that is, the generalist is sufficiently close to the initial condition that it may be found by the diffusing population of antibodies:

	\be
	\sum_{d = 0}^{d_{i \to g}} \binom{L}{d} < N
	\ee

	\indent For $L = 19$ and $N = 500$, as used in the simulation, this suggests that for $d_{i \to g} \leq 4$ the generalist may be reasonably found by diffusion.
\end{enumerate}

\indent Assuming that we are in the latter regime, we may estimate $\tau_{\text{min}}$ from $\langle d (t) \rangle$, the average distance away from the initial condition that an individual has diffused in time $t$, by solving:

\be \label{eq:taumin_definition}
\langle d( \tau_{\text{min}} ) \rangle \sim d_{i \to g}
\ee

\indent We compute $\langle d(t) \rangle $ as follows: the probability that a diffusing individual may be found at (Hamming) distance $d$ from its initial genotype is:

\be \label{eq:diffusingaway}
P(d,t) = \binom{L}{d} \, \me^{-\mu t} \sinh^d \l \f{\mu t}{L} \r \cosh^{L-d} \l \f{\mu t}{L} \r
\ee

\indent Thus,

\be
\langle d (t) \rangle = \sum_{d = 0}^{L} d \, P(d,t) = L \me^{-\mu t} \sinh \l \f{\mu t}{L} \r
\ee

Inserting into Eq. \ref{eq:taumin_definition} and solving for $\tau_{\text{min}}$:

\be
\tau_{\text{min}} = \f{L}{2\mu} \, \log \l \f{L}{L - 2 d_{i \to g}} \r  \approx \frac{d_{i \to g} }{\mu}
\ee

\noindent where the approximation is valid for $d_{i \to g}/L \ll 1$. For values used in the simulation ($L = 19$, $d_{i \to g} = 3$), we obtain $\mu \tau_{\text{min}} \approx 3.5$, which is consistent with our numerical results.

\subsubsection*{Maintaining a generalist: $\tau_{\text{max}}$}

\indent We may similarly estimate the upper bound for effective cycling, $\tau_{\text{max}}$. Supposing that an evolving population has found the generalist, it must now remain localised there. For this to happen, the environment must switch rapidly enough to prune, by purifying selection, those individuals that diffuse away from the generalist.

\indent Consider a monoclonal population of size $N$ at a generalist at time $t = 0$. For simplicity, we assume that any single mutation out of the generalist is into a specialist. This is reasonable when the number of generalists is small compared to specialists, as is the case for our numerical calculations. Then, from Eq. \ref{eq:diffusingaway}, the number of generalists remaining at time $t$ is:

\be
P(d = 0, t) = \me^{-\mu t} \cosh^L \l \f{\mu t}{L} \r
\ee

\indent Then, $\tau_{\text{max}}$ is defined as the time taken for the occupancy of the generalist to fall below 1 individual:

\be
P(d = 0, \tau_{\text{max}}) = \f{1}{N} \Rightarrow \tau_{\text{max}} = \f{L}{2\mu} \log \l \f{N^{1/L}}{2 - N^{1/L}}  \r \approx \f{1}{\mu}\log N
\ee
\noindent where in the last step we have taken the limit of large sequence length $L$.

\indent For the parameters used in the simulations ($L = 19$, $N = 500$), we have $\mu\tau_{\text{max}} \approx 8$, which is consistent with the numerical results.

\subsubsection*{Existence of an intermediate timescale}

\begin{figure}
    \centering
    \includegraphics[width=\textwidth]{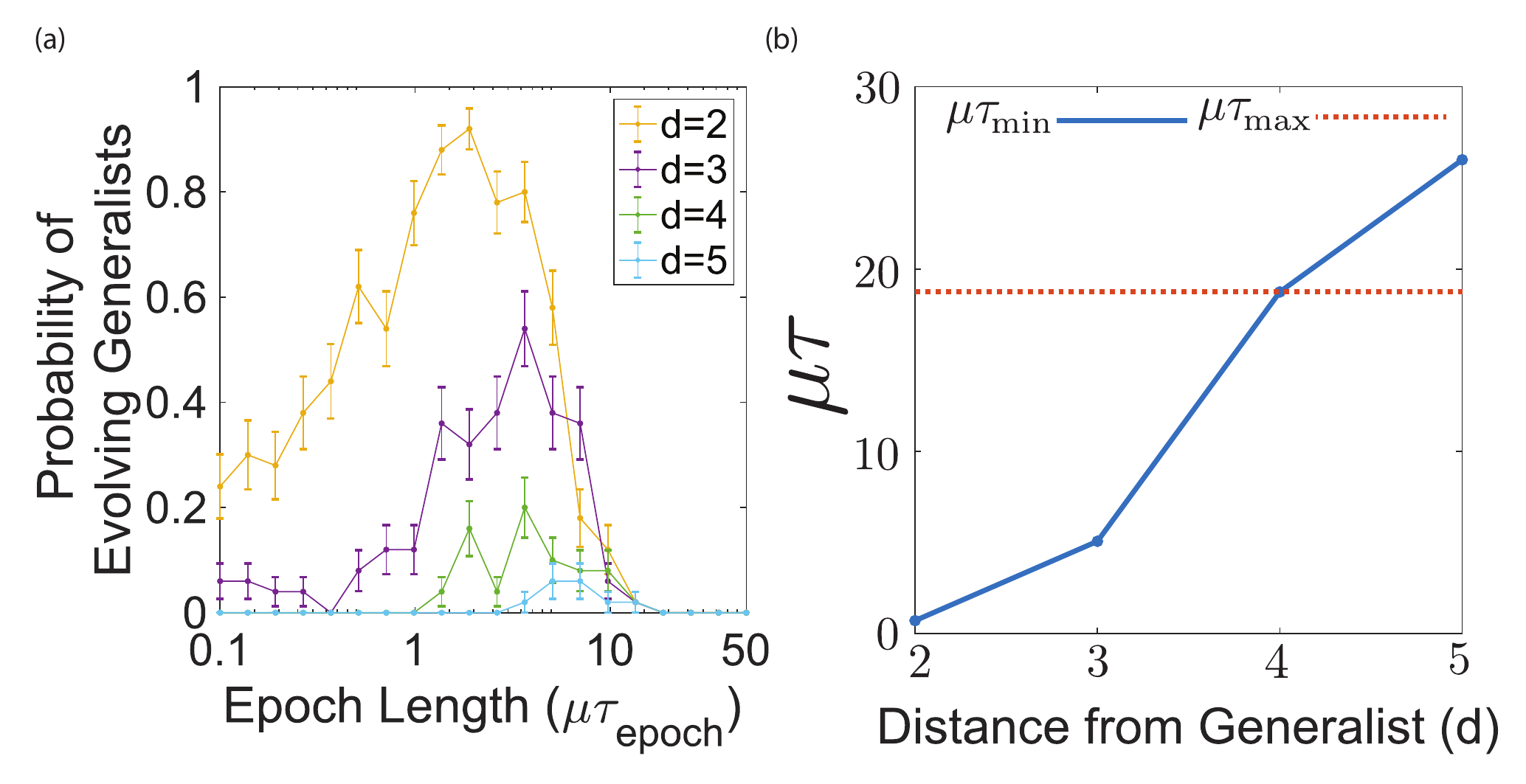}
    \caption{(a) We compute the probability of evolving generalists for varying Hamming distances from the generalist, given by $d$, using the method described in Section \ref{sec:finitepopsim}. We note that as $d$ increases, the time at which generalists begins to be evolved increases, though the time at which they probability decays remains fixed. (b) We compute $\chi_{s\rightarrow g}$ for each initial condition and take $\tau_{\text{min}}$ to be the smallest time where $\chi_{s\rightarrow g}>0.6$. We find that this rises, eventually rising above $\tau_{\text{max}}$ for large enough $d$. By construction, $\tau_{\text{max}}$ is independent of $d$.}
    \label{fig:varytaumin}
\end{figure}

Existence of an intermediate timescale $\tau_{\text{epoch}}$ that produces generalists requires $\tau_{\text{min}} < \tau_{\text{max}}$. However, the above expressions make it clear that $\tau_{\text{min}} < \tau_{\text{max}}$ only if (a) the initial condition is close enough to generalists (small $d$), (b) the fraction of generalists relative to specialists is large enough (large $L_c/L$).  

SI Fig.Fig.\ref{fig:varytaumin}a shows how the yield of generalists at intermediate timescales disappears as the initial conditions are made less favorable. This panel was constructed using the simulation method described in Section \ref{sec:finitepopsim} with variations on the initial condition for $\mathbf{x}$ such that the Hamming distance between $\mathbf{x}$ and the generalist varied by some distance $\text{d}$. SI Fig.\ref{fig:varytaumin}b shows that $\tau_{\text{min}}$ rises in this limit and exceeds $\tau_{\text{max}}$. We approximated $\tau_{\text{min}}$ by computing the smallest time for which $\chi_{s\rightarrow g}>0.6$ and $\tau_{\text{max}}$ by computing the largest time for which $\chi_{g\rightarrow g}>0.6$. 

When $\tau_{\text{min}} < \tau_{\text{max}}$ is not satisfied, there is no intermediate timescale. The time needed for generalists to specialize is shorter than the time needed to evolve generalists from specialists. In this case, the chirp protocol described below is still successful at producing generalists.

\subsection{Simultaneous presentation of antigens}

The cycling strategy explored in this paper may not be practical in the fast limit in the context of B-cell affinity maturation. A common practical alternative is vaccination with a cocktail of antigens, i.e., simultaneous exposure to multiple antigens.

Such simultaneous exposure to multiple antigens is mathematically equivalent to fast cycling of those antigens if specific microscopic assumptions about antibody-antigen interactions in germinal centers hold\cite{Wang2015}. During the affinity maturation process, folicular dendritic cells (FDCs) host antigens on their surface for B-cells to interact with, and if antibodies expressed on B-cells bind the antigens with high enough affinity, B-cells internalize those antigens. This process enables those B-cells to avoid apoptosis, and  thus proliferate and continue affinity maturation. 

There are two currently experimentally unresolved hypotheses about antigen presentation by FDCs:

1. Antibodies are fit only if they can bind \emph{ALL} presented antigens: In this hypothesis, FDCs only present a single antigen or present antigens in a spatially heterogeneous manner. Consequently, each B-cell is randomly exposed to a single antigen at the selection stage of the affinity maturation process. A B-cell must be able to bind \emph{ALL} presented antigens to survive selection.

Such simultaneous presentation is qualitatively similar to the fast cycling limit studied in this paper. When presented with such a cocktail vaccine, antibodies starting from a naive repertoire are expected to go extinct since such antibodies typically cannot bind all antigens with high affinity, as seen in the experiments of \cite{Wang2015}. 

2. Antibodies are fit if they can bind \emph{ANY} one of the presented antigens: If the FDCs present the antigens in a homogeneous manner,  each B-cell only needs to bind \emph{ANY} single one of the presented antigens to avoid apoptosis. 

In this case, specialists are fit enough to survive early rounds of selection and evolve generalists. But generalists cannot be maintained in preference over specialists unless selection pressures are fine tuned (e.g., specialists are strongly out-competed by generalists once generalists evolve, despite specialists having significant fitness to begin with).  We do not consider this model of B-cell - FDC interactions in this paper.



\subsection{Death and fast cycling}


In the fast cycling limit, $\tau_{\text{epoch}} \rightarrow 0$, the fitness of specialist antibodies is the average of their fitness in different environments; as seen Eqn. \ref{eq:hisiLandscape},  this fitness is $s(\epsilon -2)$.

As discussed above for simultaneous presentation, we only consider the case where fast cycling corresponds to hypothesis (1), where antibodies need to bind all antigens to survive. Hence, we need $s(\epsilon -2)$ to be sufficiently negative, so that a specialist population of size $N$ typically dies out before reaching the generalists in this fast cycling limit. Since the latter process takes time $\tau_{\text{min}} \sim d_{\text{i}\to\text{g}}/\mu$ as derived earlier and the initial population size is $N$, the condition for a specialist population to go extinct in the fast limit is $N \exp( s(\epsilon - 2) \tau_{\text{min}}) \sim N \exp( s(\epsilon - 2) /\mu ) < 1$. Assuming that $1 < \epsilon < 2$, we find $s > \mu \log N$ as a conservative criterion independent of $\epsilon$.

\begin{figure}
\begin{centering}
\includegraphics[width=\linewidth]{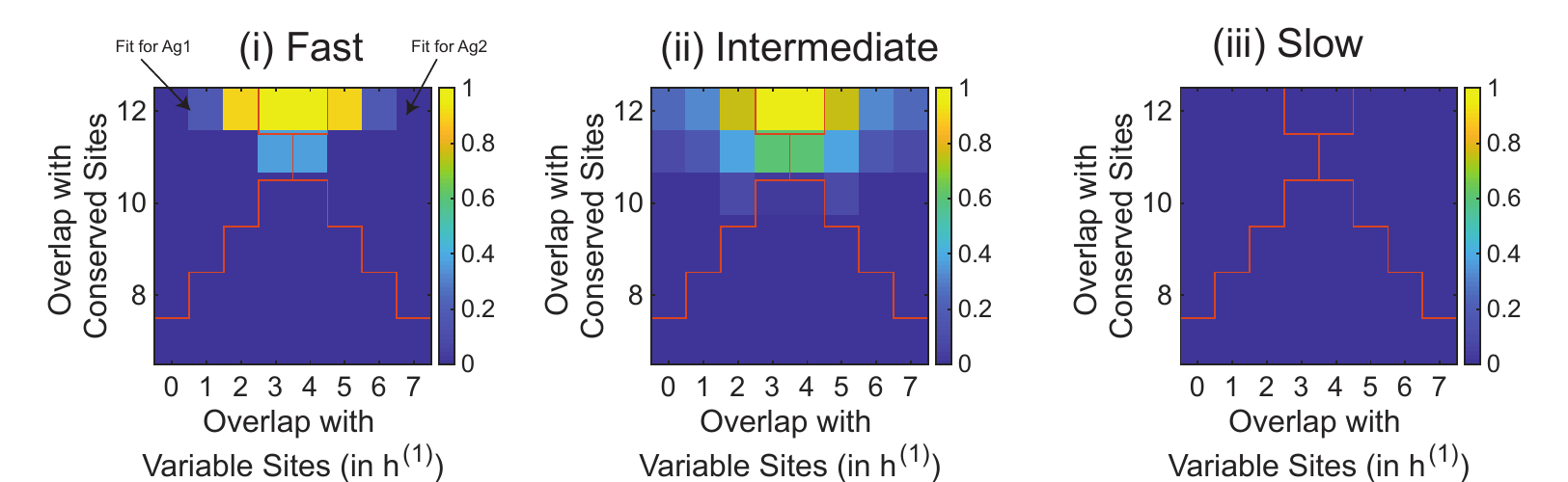}
\caption{Intermediate cycling increases the effective attractor size of generalists in models of entropically disfavored generalists. Here, the color in the color map represents the proportion of trials leading to a generalist, where yellow corresponds to all trials, while blue corresponds to no trials leading to a generalist. We initialize our population at size $N=10$ with a carrying capacity $K=500$. We maintain $L=19$, $L_{c}=12$, and $T=11$. We ran our simulations for a total of ten epochs. We ran each ordered pair of conserved and variable matches for $50$ replicates in environment $\eta=1$ and assumed symmetry over environments. (i) Fast cycling ($\mu\tau_{\text{epoch}}=0.05$) can only evolve generalists when initial conditions are already very close to generalists. (ii) Intermediate cycling ($\mu\tau_{\text{epoch}}=3$) increases the number of viable initial conditions. (iii) Slow cycling ($\mu\tau_{\text{epoch}}=15$) does not lead to generalists for any initial condition since this limit is unable to maintain generalists.}
\label{attract_size}
\end{centering}
\end{figure}

\subsection{Dependence on initial repertoire}
In the main paper, we initialize our population from a genotype that exactly binds the antigen characterized by genotype $\mathbf{h}^{(1)}$. In SI Fig.\ref{attract_size}, we show that not only does intermediate cycling increase the likelihood of evolving generalists from a given initial condition, it also increases the number of initial conditions (e.g., initial B-cell repertoire) that can lead to generalists. Hence, intermediate cycling increases the effective attractor size of the fitness peak associated with a generalist. In SI Fig.\ref{attract_size}, we consider cycling fast ($\mu\tau_{\text{epoch}}=0.05<\mu\tau_{\text{min}}$), cycling at an intermediate rate ($\mu\tau_{\text{epoch}}=3$), and cycling slowly ($\mu\tau_{\text{epoch}}=15>\tau_{\text{max}}$). Blue regions in the heatmap correspond to initial conditions that led to few surviving populations after ten epochs. Yellow regions in the heatmap correspond to initial conditions that led to many surviving populations after ten epochs. We ran each ordered pair of conserved matches and variable matches for $50$ replicates in environment $\eta=1$ and symmetrized over environments.

\subsection{Chirp protocol}

\textbf{Trade-off in fixed frequency cycling:}
The anticorrelated behavior of $\chi_{\text{s}\to\text{g}}$ and $\chi_{\text{g}\to\text{g}}$ is indicative of a trade-off between evolving generalists and maintaining them in the population. 

We first assess this by only considering simulation runs used in Fig.2b that did not result in extinction and computed the number of generalists at the end of such simulations. This number is plotted in Fig.3a. We note that as epoch length increases, the number of generalists remaining in the population decreases, but the probability of a population evolving a single generalist increases. 

To illustrate this point we compute two quantities for each simulation for a given $\tau_{\text{epoch}}$: 
\begin{itemize}
    \item The number of generalists (if non-zero): The number of generalists is simply the average number of generalists in the population for simulations where the population survived an evolutionary run. This is plotted on the y-axis of Fig.3c.
    \item The probability of a surviving population: The proportion of trials for a given $\tau_{\text{epoch}}$ that a population does not go extinct during the evolutionary run. This is plotted on the x-axis of Fig.3c.
\end{itemize}
By plotting these two quantities against each other, as in Fig.3c in black dots, we observe a tradeoff front. 

\textbf{Chirp Cycling Breaks the Trade-off:}
This trade-off leads us to proposing a `chirp' protocol. In a `chirp' cycling protocol, we decrease the length of the epoch using a multiplicative factor after each cycle, enabling us to take advantage of high probability of population survival (favored by slow cycling) and still obtain high yield (favored by fast cycling). We update $\tau_{\text{epoch}}$ according to the following rule:
\begin{gather}
    \tau_{\text{epoch}} \leftarrow k \tau_{\text{epoch}}
\end{gather}
where $k$ is some number smaller than $1$. We continue evolving the population until $\tau_{\text{epoch}}<<\lambda$. Plotted in Fig.3b is a time trace of the population size and the fraction of the population that is of a generalist genotype. Generalist fraction is $\frac{\text{Number of Generalists}}{\text{Population Size}}$. We note that as the length of each epoch decreases, the generalist fraction decreases less in time, until eventually, it remains stabilized at $\approx1$. Additionally, fluctuations in population size are suppressed. 
We ran the chirp protocol for $25$ replicates and computed the number of generalists at the end of each run, if the population survived the run, and the probability that the population survived a chirp protocol. Plotting this in Fig.3c. demonstrates that the tradeoff boundary has been broken.

\textbf{Chirped Cycling Evolves Generalists when $\tau_{\text{max}}<\tau_{\text{min}}$}
By implementing the same chirped strategy for initial conditions where $\tau_{\text{max}}<\tau_{\text{min}}$, we find that we have high generalist yield at rates match are near the maximum probability of evolving generalists of the fixed frequency evolutionary runs.

\section{Generalists Separated by Barriers}\label{sec:GenSep}

\subsection{Model}







We model the fitness landscape of an antibody binding to an antigen with multiple epitopes through a phenomenological construction, inspired by Hopfield's spin glass landscape. 

Consider an antigen $\eta$ with $P_{\eta}$ epitopes (i.e., sets of residues on the antigen that form binding locations for antibodies). Suppose that for each epitope, an antibody with sequence $\mathbf{h}^{(\eta)}_{\alpha}$ of length $L$ with entries $\pm1$ binds with high affinity (with $\alpha \in \{1, ..., P_{\eta} \}$ indexing epitopes). Then, the overall binding affinity to antigen $\eta$ of any  antibody with sequence $\mathbf{x}$, of length $L$ with entries $\pm1$ is taken to be:

\be
F^{(\eta)}(\mathbf{x}) = \sum_{\alpha} \kappa^{(\eta)}_{\alpha} \, \l \mathbf{h}^{(\eta)}_{\alpha} \cdot \mathbf{x} \r^2
\ee 

\indent This construction naturally produces islands of high fitness around the epitope-binding antibodies $\mathbf{h}^{(\eta)}$, separated by regions of low fitness. Here, $\kappa^{(\eta)}_{\alpha}$ is the binding affinity of the ideal antibody $\mathbf{h}^{(\eta)}$ to its cognate epitope. We shall take epitope $\alpha = 1$ to be present for all antigens, thus defining the generalist. In keeping with the assumption that the generalist is less fit in any landscape, we take its binding affinity $\kappa^{(1)}_1 = \kappa^{(2)}_1 = 0.08$ and all other $\kappa^{(\eta)}_{\alpha} = 0.1$.


\textbf{Weighting the various solutions:} We impose that the height of the fitness peak associated with the generalist is lower than the peaks of the specialist ($\frac{\kappa^{(\eta)}_1}{\kappa^{(\eta)}_{\alpha\neq1}}=0.8$), reflecting fitness costs associated with being a generalist relative to specialists in a fixed environment.

\subsection{Population Simulations}\label{sec:Moran}

We simulate a population of antibodies evolving in these landscapes by implementing a Moran process\cite{moran1958} with three events:
\begin{itemize}
    \item \textbf{Environment shifts} with a deterministic rate, $\frac{1}{\tau_{\text{epoch}}}$
    \item \textbf{Mutation} with a rate, $\mu$ per individual, where a single site on $\mathbf{x}$ is bit-flipped
    \item \textbf{Reproduction} with a rate, $\lambda$ per individual, where an individual is selected from the population with probability proportional to $\exp(F^{(\eta)})$ to be duplicated and another individual from the population to be removed with uniform probability
\end{itemize}
and a population size of $N$. 

In our population simulations, we impose the following values for our simulation parameters. We fix the total number of epitopes for each landscape, $P_{\eta}=11$ across all $\eta$, keeping just one generalist. We impose sequence length to be $L=100$, generating each optimal epitope-binding antibody randomly. We initialize our simulations from a random monoclonal initial condition of size $N=100$ unless otherwise specified. We impose a per individual mutation rate of $\mu=0.25$ and a reproduction rate of $\lambda=1$. 

\subsection{Fig.4b: Resonance Peak in Generalist Discovery as a function of $\tau_{\text{epoch}}$}
We ran our simulation for $50$ replicates from random monoclonal initial conditions of size $N=100$ with sequences of length $L=100$. We initialize the landscapes with $10$ specialist antibodies and $1$ generalist antibody (i.e. $P_{\eta}=11$ for all $\eta$). Our simulations were run for a total of $100\text{ }\tau_{\text{epoch}}\text{s}$. We swept over $\tau_{\text{epoch}}$s. We note that during the simulations, regardless of the frequency of environmental shifts, the population remains tightly clustered as it evolves in time. This behavior corresponds to evolution in the strong selection and weak mutation limit. We consider a generalist to have been discovered if, after a run, there exists at least one individual whose overlap with the generalist antibody is $90\%$. Using these simulations, we demonstrate that an intermediate regime of switching enhances the discovery of generalists. We plot the results of these simulations in Fig. 4b. 

\begin{figure}
    \centering
    \includegraphics[width=0.8\linewidth]{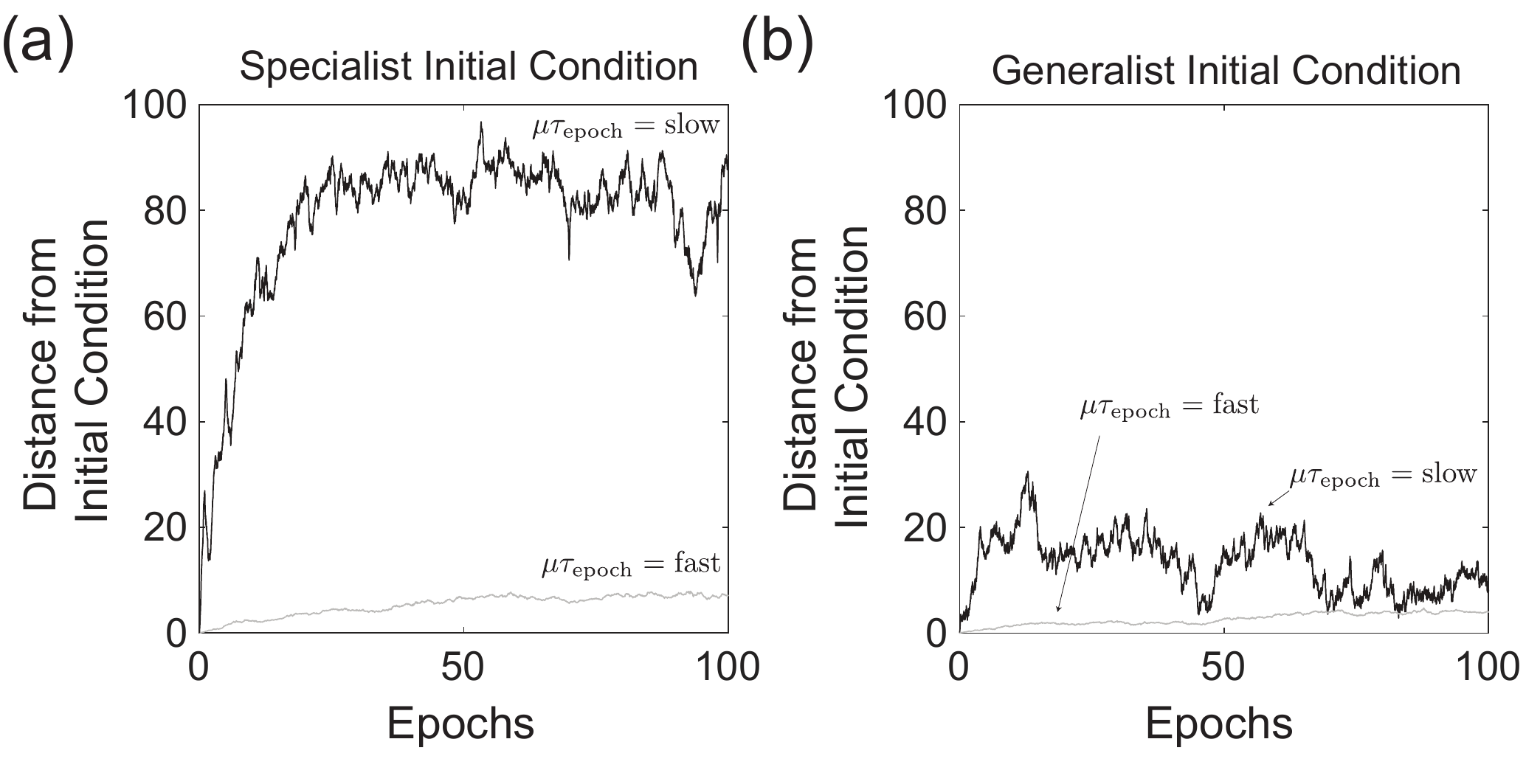}
    \caption{Specialist populations evolve significantly through sequence space for intermediate timescale cycling but not fast cycling; generalists do not evolve significantly for any timescale cycling. (a) An initially specialist population does not evolve away from the initial genotype for fast cycling. However, with slower cycling, the population evolves to a significantly different genotype(s). (b) An initially-generalist population does not significantly evolve away from the initial genotype for fast or slow cycling.}
    \label{fig:hopfieldtimetrace}
\end{figure}

\textbf{Time traces of the population from its initial condition:} To identify the reason for this resonant peak, we run the simulation for a $100$ cycles and plot in SI Fig.\ref{fig:hopfieldtimetrace} how far the population is from its initial condition in hamming distance after a fixed number of epochs We initialize a population at a specialist antibody and a generalist antibody, comparing the behavior for fast cycling and slow cycling. We show that for generalist initial conditions, regardless of cycling rate, the population remains in the generalist. There is some fluctuation out but the population returns to the generalists often. Given enough time, the population will escape though this is a slow process. However, for specialist initial condition, slow cycling enables the population to escape its initial condition, while fast cycling does not allow such escape.
\subsection{Timescale Analysis}
\textbf{Computing the minimum epoch length, $\tau_{\text{min}}$:} 

Cycling benefits evolution of generalists in fitness landscapes with barriers by enabling the population to escape specialist peaks in one landscape by evolving subject to a different fitness landscape.  To escape from a specialist peak in environment 1, the population must accrue enough mutations when subject to environment 2, such that when environment 1 returns again, the population is not likely to return to the original specialist peak. 

We first consider the strong selection limit $s N \gg 1$. Consider a monoclonal population at a specialist peak $\mathbf{h^{(1)}}_\alpha$ in fitness landscape $F^{(1)}$. When such a population is now subject to landscape $F^{(2)}$ for a time $\tau_{\text{epoch}}$, $\mathbf{h^{(1)}}_\alpha$ serves as an initial condition of typical low fitness and will evolve towards a fitness peak $\mathbf{h^{(2)}}_\beta$ in $F^{(2)}$.  If we switch back to $F^{(1)}$ after a sufficiently long time, the population genotype $\mathbf{x}$ will be  sufficiently mutated compared to $\mathbf{h^{(1)}}_\alpha$ that the population will likely fix to an alternative fitness peak $\mathbf{h^{(1)}}_\beta$ in $F^{(1)}$. Let us assume that the number of such mutations needed is $d_{12}$. 

Since the population in genotype $\mathbf{h^{(1)}}_\alpha$ is typically of low fitness in landscape $F^{(2)}$, most mutations are beneficial. Then, in the strong selection limit, the time needed to acquire $d_{12}$ beneficial mutations is by the mutation rate,
 
$$\tau_{\text{min}} \approx \frac{d_{12}}{\mu}.$$
Fig.

This minimal number of mutations $d_{12}$ to escape the `attractor basin' of a fitness peak $\mathbf{h^{(1)}}_\alpha$ is model dependent.  
- $d_{12}$ depends on the size of the attractor basin around $\mathbf{h^{(1)}}_\alpha$ and the correlations between $F^{(1)}$ and $F^{(2)}$. In our Hopfield-inspired model of fitness landscapes $F^{(\eta)}$, if the fitness peaks are randomly distributed in a sequence space of length $L$, then the empirical value of $d_{12} \sim  \frac{1}{4}L$.  In real fitness landscapes, this distance $d_{12}$ can vary widely for different specialists which can have attractor regions of different size.



\textbf{Computing the maximum epoch length, $\tau_{\text{max}}$:} 

Unnecessarily long times $\tau_{\text{epoch}} > \tau_{\text{min}}$ spent in each environment is counter-productive. To see this, note that specialists are most likely to evolve to generalists in a short duration of time after an environmental switch. Any extra time spent $\tau_{\text{epoch}} > \tau_{\text{min}}$ in the same environment is simply `dead time' that does not increase the yield of generalists further. Hence the effective rate of evolving generalists from specialists falls as $1/\tau_{\text{epoch}}$ for  $\tau_{\text{epoch}} > \tau_{\text{min}}$. 

Meanwhile, existing generalists can specialize again. Let the rate of this process be $r_{{g \to s}}$. The yield of generalists is  reduced when this escape rate $r_{g \to s}$ from generalists to specialists is larger than the switching-induced rate from specialists to generalists $1/\tau_{\text{epoch}}$. Hence, $\tau_{\text{max}} \sim 1/r_{g \to s}$.



The rate $r_{g \to s}$ at which generalists specialize is easily estimated since the fitness landscape does not change in time for sequences near the generalist. Hence this process is the well-studied process of an asexual population crossing a fitness valley by picking up a sequence of deleterious mutations. This process has been studied in numerous regimes with different assumptions about population sizes, selection pressure \cite{Weissman2009, Van_Nimwegen2000, Jain2007}.  Here, we assume strong selection and weak mutation, allowing us to use the simple result $r_{gs} \sim \mu \, e^{- N \Delta F_g}$ result obtained from the analogy of statistical physics and population dynamics; population size $N$ plays the role of temperature and $\Delta F_{g}$, the fitness difference between the generalist peak and the fitness valley, plays the role of an energy barrier. Hence, $$\tau_{\text{max}} \approx \frac{\exp(N \Delta F_{g})}{\mu}.$$

Real populations can often violate these assumptions; in that case, any other relevant result\cite{Jain2007,Van_Nimwegen2000,Weissman2009} for valley crossing rates can be used in place of $r_{g \to s}$. 




\subsection{Fig.4c: Transitions Amongst Specialists and Generalists: $\chi_{\text{s}\to\text{g}}$ and $\chi_{\text{g}\to\text{g}}$}

The presence of a resonant peak in Fig. 4b is suggestive an underlying tension between discovering the generalist and escaping the generalist, similar to that in the earlier model of entropically disfavored generalists. As such, we re-introduce the quantities $\chi_{\text{s}\to\text{g}}$ and $\chi_{\text{g}\to\text{g}}$: 

\begin{itemize}
    \item $\chi_{\text{s} \to \text{g}}$ is the proportion of trials initialized from a monoclonal random initial condition that evolve a generalist (i.e. a single member of the population matches $90\%$ of the generalist) antibody emerges within $30$ epochs of an evolutionary run for a given $\tau_{\text{epoch}}$
    \item $\chi_{\text{g} \to \text{g}}$ is the proportion of the population, when started from an initially generalist population (i.e., all $\mathbf{x}=\mathbf{h}^{(\eta)}_{1}$), remains $90\%$ overlapped with the generalist that remain in a generalist (i.e. maintains $90\%$ overlap with $\mathbf{h}^{(\eta)}_{1}$ on average) after $100$ epochs for a given $\tau_{\text{epoch}}$
\end{itemize}

We chose $100$ epochs in the definitions above as beyond $100$ epochs, marginal changes in both values decreases. Plots for $\chi_{\text{g}\to\text{g}}$ and $\chi_{\text{s}\to{g}}$ are shown in Fig. 4c and illustrate the same behavior as in the entropically disfavored models.

\subsection{Cycling-induced variance and correlations between environments}

To illustrate how cycling enables the discovery of generalists, we consider population trajectories during cycling. We consider the impact of the initial condition of the population on these trajectories and the impact of the correlation structure between the different environments. To this end, we introduce a measure of correlation and introduce a new simulation to capture the effective behavior of the population.

\subsubsection{Definition of Correlations between Environments}
 We measure the correlation between landscapes, denoted $\langle F^{(1)} | F^{(2)} \rangle_{s}$ using the following equation:
\begin{gather}
    \langle F^{(1)} | F^{(2)} \rangle_{s} \equiv \frac{c(\mathbf{h}^{(1)}, \mathbf{h}^{(1)})}{c(\mathbf{h}^{(1)}, \mathbf{h}^{(2)})c(\mathbf{h}^{(2)}, \mathbf{h}^{(2)})}
\end{gather}

where we have defined the function $c(\mathbf{h}^{(1)}, \mathbf{h}^{(2)})=\sum_{\alpha_{1}, \alpha_{2}=2}^{P_{1},P_{2}} \mathbf{h}^{(1)}_{\alpha_{1}} \cdot \mathbf{h}^{(2)}_{\alpha_{2}} / (L\sqrt{P_{1}P_{2}})$. 

This measure of correlation is high if the specialist genotypes for different antigens are highly similar in pairs; e.g., if each specialist for antigen 1 is similar to a specialist for antigen 2. As seen below, a high correlation by this measure implies that specialist antibodies do not evolve significantly due to cycling and thus generalists are not easily evolved. 

Note that this measure is normalized so the measure is unaffected by the diversity $c(\mathbf{h}^{(\eta)}, \mathbf{h}^{(\eta)})$ of specialist genotypes for a single antigen $\eta$.


\subsubsection{Modeling Population Trajectories with Single Walkers}\label{Metropolis}

To measure the role of cycling between landscapes and the correlation structure of the landscape, we studied the dynamics of single walkers. This is justified as the population is shown to be roughly monoclonal in its evolutionary trajectories. Single walkers were simulated via the well-known Metropolis-Hasting algorithm\cite{Metropolis1953}. We preserve definitions of $\mathbf{x}$, $F^{(\eta)}(\mathbf{x})$, and all related quantities from before. The process is as follows: 

\begin{itemize}
    \item Randomly select a single site to mutate to create new variant $\mathbf{x}'$ from original $\mathbf{x}$
    \item Compute fitness of new variant
    \item Accept new variant with probability $\exp(\beta(F^{(\eta)}(\mathbf{x}')-F^{(\eta)}(\mathbf{x})))$ and repeat.
\end{itemize}

Because of the differences between single walker dynamics and population dynamics, we include an overall scale for the landscapes, $\beta$. $\beta$ is chosen to be $\beta=4$.

\subsubsection{Fig.4f: Cycling-Induced stochasticity}
We begin by considering antigens with uncorrelated specialists (ie, $\langle F^{(1)} | F^{(2)} \rangle_{s}\approx 0$). Starting from two initial conditions, a generalist antibody and a specialist antibody for antigen $\eta=1$, we evolve the walker for $k$ proposals in the presence of antigen $2$, and then allowed the walker enough proposals to relax to a stable solution in the presence of antigen $1$. By computing the final state for $20$ different walkers in a given landscape, and averaging over $20$ random landscapes, we can compute the variance in the final positions of the walkers. This is accomplished by computing the average pairwise distance between walkers in the same landscape, and then averaging over landscapes. To demonstrate the importance of the number of proposals, $k$, which is serving as a proxy for $\tau_{epoch}$, we swept over $k$. The result is plotted in Fig. 4f. 

We see that when starting from a specialist initial condition, cycling-induced variance rises when $\tau_{epoch}$ is sufficiently large. Generalists, as predicted, are unaffected by cycling, as those genotypes are fit in both environments.

\subsubsection{Fig.4g: Impact of Correlation Structures Between Cycled Environments}
We then repeated the same simulations as above with increasing correlation structure between the landscapes. We enforced that $k=250$ was large enough to ensure high stochasticity in uncorrelated environments. We see that as correlation between the landscapes rises, cycling-induced stochasticity decreases. The results are plotted in Fig. 4g. 

\subsection{More than 2 Antigens}

Throughout the main text, we only consider the evolution of generalists in the presence of two antigens. Here, we demonstrate that when we increase the number of antigens, discovery of the generalist becomes easier. We demonstrate this in the Moran simulations by increasing the number of randomly constructed landscapes with a single generalist peak shared across the landscapes. In order to probe a  dynamic range of antigen number, we weight the generalist to be smaller than in previous trials, setting $\kappa^{(\eta)}_{1}=0.07$, rather than $0.08$. We ensure each antigen is presented at least once, cycling for at least $30$ epochs. We maintain the parameters used before. 

\begin{figure}
\begin{centering}
\includegraphics[width=75mm]{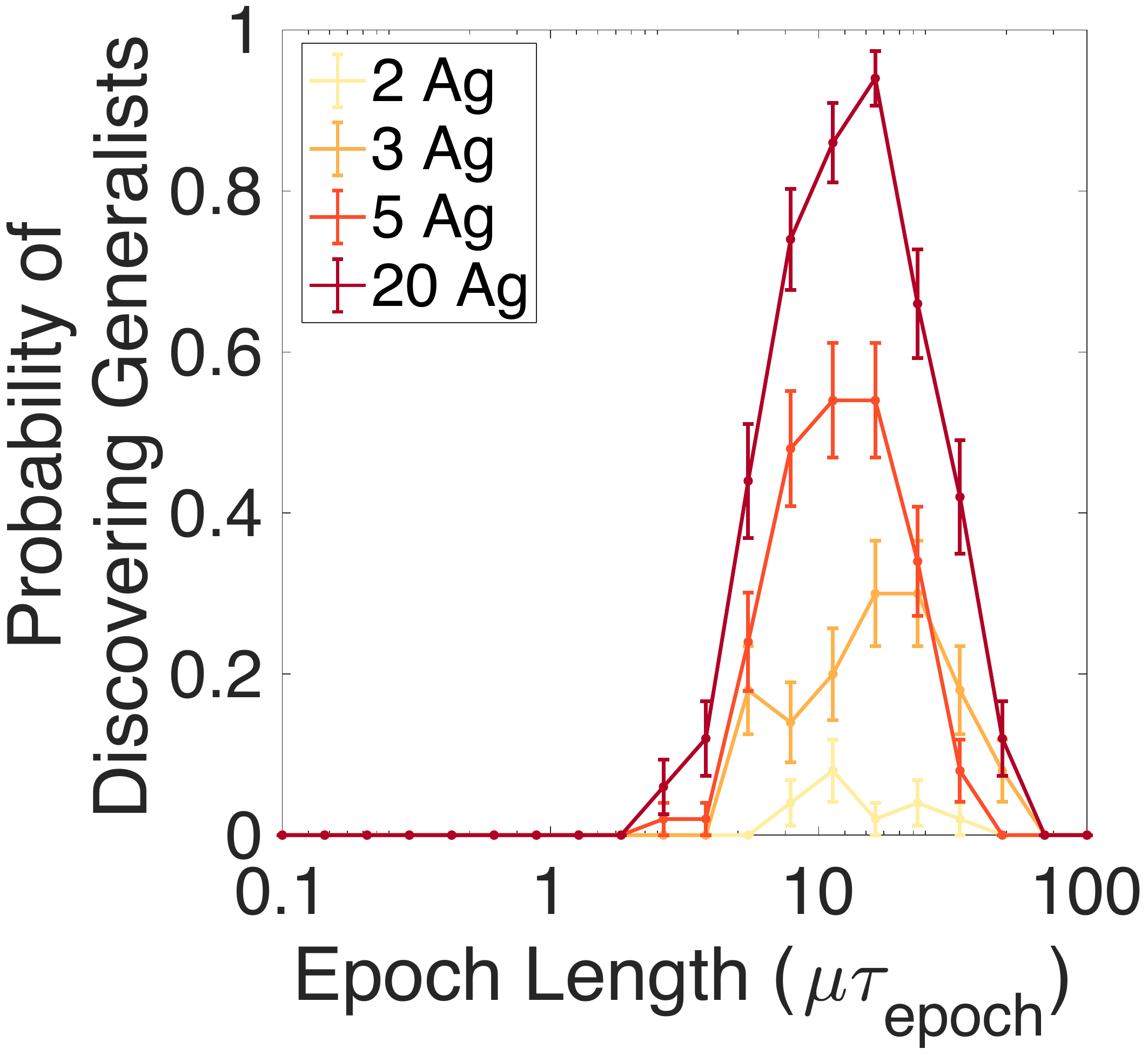}
\caption{An increased number of distinct antigens makes it easier to evolve generalists through cycling. We increased the number of distinct antigens in stepsfrom $2$ through $20$, each with $10$ randomly chosen specialist epitopes, and one generalist epitope common to all of them. We cycle between these landscapes using the Moran simulation described in Section \ref{sec:Moran} for 30 $\tau_{\text{epoch}}$. We consider the evolutionary run to have evolved a generalist if at least one antibody has an overlap of $90\%$ with the generalist. We find that increasing the number presented increases the likelihood that generalist genotypes are discovered.} 
\label{fig:varyag}
\end{centering}
\end{figure}

We find that increasing the number of antigens increases the discovery-likelihood of generalists, as plotted in SI Fig.\ref{fig:varyag}. We interpret these results as due to effectively reduced correlation between landscapes since correlations shared across a subset of the antigens may not be shared across all antigens. For example, the population can settle into a limit cycle between specialists with only 3 antigens but evolution in the presence of other antigens allows escape from such cycles. As a result, we increase the rate of evolution from specialists to generalists without enhancing the reverse process. 

\begin{figure}
\begin{centering}
\includegraphics[width=75mm]{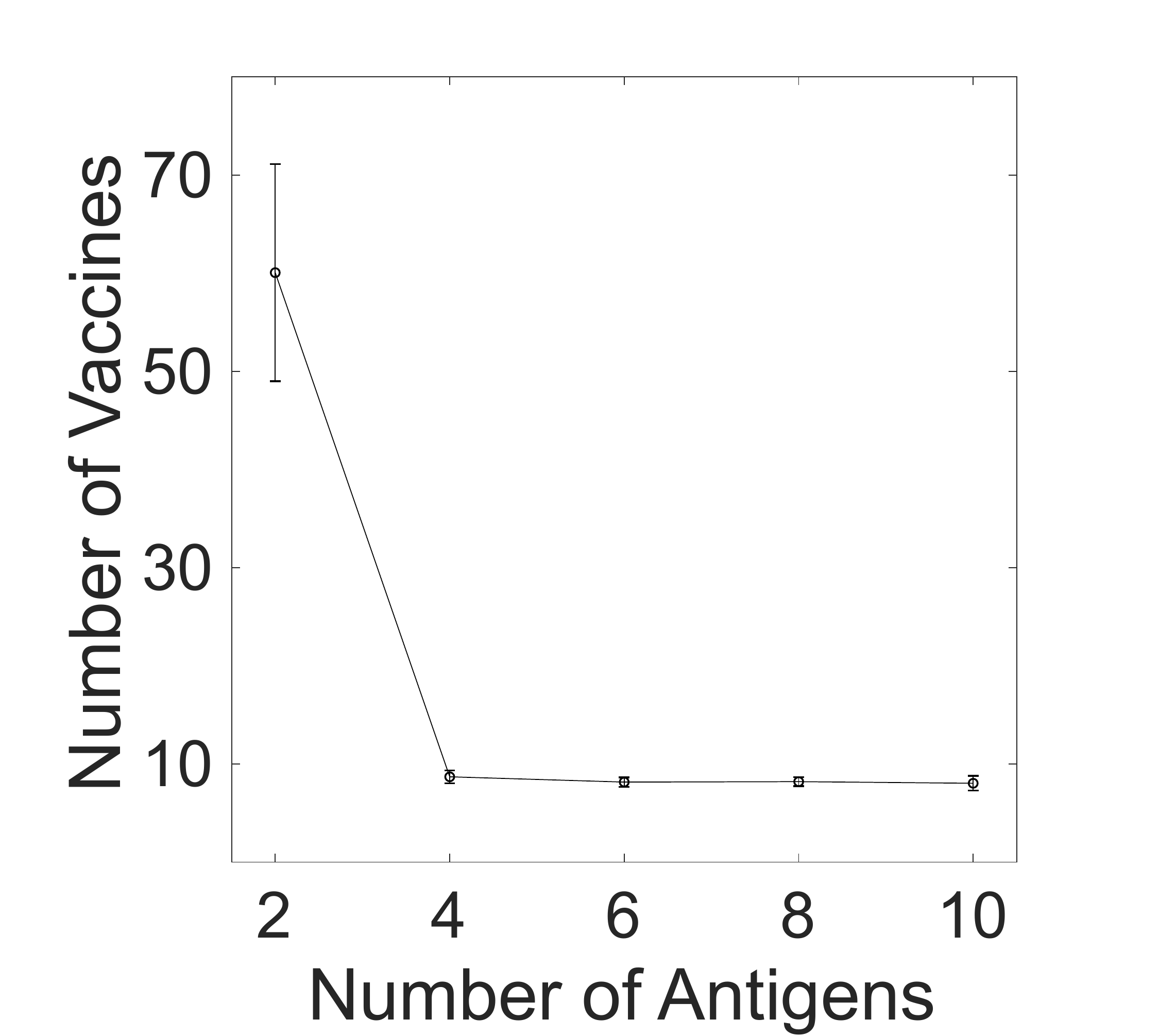}
\caption{The number of vaccine shots required to evolve generalists is reduced if distinct antigens are used. We repeat the simulation described in SI Figure \ref{fig:varyag}, except we run until a generalist is discovered for a fixed $\mu\tau_{\text{epoch}}=40$. We compute the average first arrival time across $50$ replicates and the standard error. We find that increasing the number of distinct antigens used decreases the number of doses needed up. } 
\label{fig:agfirstarrival}
\end{centering}
\end{figure}

We further consider the number of cycles needed for the first generalist to appear as a function of the number of antigens. This can be interpreted as the number of vaccine doses needed when using a particular number of strains in the vaccine course. We probe this by running Moran dynamics at a fixed $\mu\tau_{\text{epoch}}=40$ with $\kappa_{1}^{(\eta)}=0.08$, as in other simulations, for as many epochs as needed to discover a generalist. We run $50$ replicates of this simulation, reporting the average time at which a generalist first appeared across those replicates. Our choice of epoch length is the epoch length at probability of evolving generalists appears to maximize. We find that the number of vaccine doses decreases as the number of antigen strains increases, as plotted in \ref{fig:agfirstarrival}.

\section{HIV Antibody Data}
The success of our proposed cycling strategy depends on specific assumptions about correlations between different antigens. In particular, antigens need to be sufficiently correlated in that the same generalist antibodies can bind them (e.g., the antigens share a epitope). And yet antigens need to be sufficiently \emph{uncorrelated}: i.e., specialist antibodies that bind different antigens must be sufficiently distinct as measured by $\langle F^{(1)} | F^{(2)} \rangle_{s}$ (e.g., the specialist epitopes on antigens must be sufficiently distinct). 

We sought to test whether these correlation conditions are met by antibodies evolved in response to real HIV strains.

\subsection{Antibody sequences and Binding Affinity Data}

Several works have studied viable antibodies from individuals afflicted with different strains of HIV\cite{Bonsignori2016, Gao2014, Liao2013}. These works sequenced viable antibodies, studied their binding affinities to different strains, and proposed intermediate antibodies in between the germline antibody and the discovered broadly neutralizing antibody. They evaluated the binding affinities of each of these antibodies using the ELISA assay. The binding affinity data is presented in SI Table \ref{table:abdata}. The mutational distance between each antibody is given in SI Table \ref{table:hamdistdata}. 

\textbf{Antibody Sequence Data:} Two classes of antibodies are presented here: mature antibodies observed in patients during their course with HIV and antibody sequences inferred to be \cite{Bonsignori2016, Gao2014, Liao2013} intermediate between the germline and the mature broadly neutralizing antibody. The natural antibodies appear with the prefix 'CH', and the inferred antibodies, which were synthesized, appear with the prefix 'IA'. 

\textbf{Antibody Binding Data:} The binding affinity of each antibody to two different strains of HIV, 31D8gp120/293F and 11D8gp120, is evaluated using the ELISA assay. Particular values for binding are presented in table \ref{table:abdata}. We impose a cutoff of 10 $\log(\text{AUC})$ to indicate when an antibody has bound an HIV strain. By this rule,  
\begin{itemize}
    \item 31D8gp120/293F is bound by antibodies IA2, IA3, CH105, and CH103. 
    \item 11D8gp120 is bound by antibodies CH186, CH187, CH200, and CH103.
\end{itemize}

\subsection{Constructing landscapes $F^{(1)}$ and $F^{(2)}$}\label{sec:realdatland}
Let $F^{(1)}$ and $F^{(2)}$ define the fitness landscape of antibody space corresponding to 31D8gp120/293F and 11D8gp120 respectively. In each landscape, the experimentally discovered and synthetically produced antibodies will define disconnected neighborhoods of antibodies that are fit for that landscape. 

We begin constructing these landscapes by converting the sequence of each antibody into a binary vector with entries $\pm1$, noting that each antibody is length $L=121$. We accomplish this randomly generating a binary vector with entries $\pm1$ of length $L$ to represent the unmutated common ancestor. Then, using the sequence data given by \cite{Bonsignori2016, Liao2013, Gao2014}, we determine where each antibody differs from the unmutated common ancestor and introduce a binary spin vector for each that preserve the differences from the unmutated common ancestor as presented in the real data. We define the sequences associated with $F^{(\eta)}$ as $\mathbf{h}^{(\eta)}_{\alpha}$. We set $\mathbf{h}^{(1)}_{1}=\mathbf{h}^{(2)}_{1}$ to the sequence of the generalist antibody CH103.

We take the binding affinity of an antibody, represented by $\mathbf{x}$ with entries $\pm1$ and length $L$, to be $E^{(\eta)}(\mathbf{x})=\sum_{\alpha}(\mathbf{h}^{(\eta)}_{\alpha} \cdot \mathbf{x})^n$. Here, $n$ controls the size and depth of each well. Here, we take $n=4$.

The fitness of each antibody is then given as:
\begin{gather}
    F^{(\eta)}(\mathbf{x})=\frac{1}{1+\exp(s(E^{(\eta)}(\mathbf{x}) - L^{n})}\\
\end{gather}
$s$ defines the overall scale of the selection pressure, as before. Here, we take $s=1$. The function thresholds the amount of fitness benefit an antibody can obtain from matching a correct sequence, so it features both degeneracy of neighboring genotypes and disconnected islands. 

\begin{table}
\begin{center}
\begin{tabular}{ |c||c|c|c|c|c|c|c|  }
 \hline

 HIV strain, Ab  & CH105 & CH186 & CH187 & CH200 & IA2 & IA3 & CH103\\
 \hline
 \hline
 31D8gp120/293F   &13.52    &1.13 & 0.00 & 5.80 & 13.34 & 13.01 & 13.63\\
 \hline
 11D8gp120 &8.97     & 13.59   &10.21  &10.92 & 9.12 & 6.82 & 10.92\\

 \hline
\end{tabular}

\caption{Binding affinity of different antibodies (columns) to two different HIV strains (rows), measured via the ELISA assay (units of the logarithm of the area under the curve (logAUC) of the absorbance of the sample)\cite{Bonsignori2016, Gao2014, Liao2013}. Higher values reflect stronger affinity. We consider an antibody to be a specialist for a strain using a cutoff of 10 logAUC. Note that only CH103 is a generalist in this dataset.}
\label{table:abdata}
\end{center}
\end{table}

\begin{table}
\begin{center}
\begin{tabular}{ |c||c|c|c|c|c|c|c|c|  }
 \hline

 Antibody & UCA & CH105 & CH186 & CH187 & CH200 & IA2 & IA3 & CH103\\
 \hline
 UCA   & 0 & 27 & 8 & 16 & 20 & 27 & 19 & 28  \\
 \hline
 CH105 & 27 & 0 & 25 & 24 & 38 & 21 & 9 & 24  \\
 \hline
 CH186 & 8 & 25 & 0 & 11 & 20 & 25 & 25 & 26 \\
 \hline
 CH187 & 16 & 24 & 11 & 0 & 27 & 23 & 19 & 24\\
 \hline
 CH200 & 20 & 38 & 20 & 27 & 0 & 37 & 31 & 38\\
 \hline
 IA2   & 27 & 21 & 25 & 23 & 37 & 0 & 15 & 4 \\
 \hline
 IA3   & 19 & 9 & 25 & 19 & 31 & 15 & 0 & 19\\
 \hline
 CH103 & 28 & 24 & 26 & 24 & 38 & 4 & 19 & 0\\
 \hline 
 
\end{tabular}
\caption{Mutational distances (Hamming Distance) between antibody sequences for antibodies observed in an HIV patient who eventually developed bnAbs. Sequences for these antibodies are found in \cite{Bonsignori2016, Liao2013, Gao2014}. Using the raw sequence data and the binding energy presented in \ref{table:abdata}, we can construct fitness landscapes $F^{(1)}$ and $F^{(2)}$ with fitness peaks that reflect these mutational distances.}
\label{table:hamdistdata}
\end{center}
\end{table}

\subsection{Simulations}

We simulated evolution using the technique described in Section \ref{Metropolis}. Given that mutation rates in B-cells undergoing somatic hypermutation are taken to be $10^{-3}$ per base pair per division\cite{Wang2015}, we choose our epoch length to be long enough that the population accumulates $100$ mutations. This corresponds to an epoch length that allows $800$ total divisions. The initial condition for these simulations was set to be the unmutated common ancestor (UCA). We note that if the population is not started from the UCA, the simulation fails to find successful antibodies.

\subsection{Shuffled assignment}

We earlier demonstrated that the correlation structure of the landscape impacted the ability of the landscape to effectively cycle its way to a generalist. Here, we find that 

\begin{equation}
    \langle F^{(1)}|F^{(2)}\rangle_{s} = 0.43
\end{equation}

which reflects the distance between specialist sequences for the two strains in the data of \cite{Bonsignori2016, Gao2014, Liao2013}  (see Table \ref{table:hamdistdata}). With such low correlations between specialists, the simulations discover generalists around $60\%$ of the time when cycling in this landscape, as shown in Fig. 5b. 

 To understand how increasing the correlation structure can impact generalist discovery in real data, we artificially shuffled the antibody binding data. In particular, we treated CH186 as a specialist antibody for 11D8gp120 and CH105 as a specialist antibody for 31D8gp120/293F and then followed the same construction of landscapes described in \ref{sec:realdatland}. In the new constrution, we find that the two things are substantially more correlated.
\begin{equation}
     \langle F^{(1)}|F^{(2)} \rangle_{s} = 0.78
\end{equation}
Then, after running simulations with changing environments, we find that recovery rates of the generalist drops significantly, as shown in Fig. 5b. 

\bibliographystyle{test}
\bibliography{IntermediateTimescales.bib}